\documentclass[]{aa}
\usepackage{xcolor}
\usepackage{graphicx}
\usepackage{txfonts}
\begin{document} 

   \title{Directly tracing the vertical stratification of molecules in protoplanetary disks}

   \authorrunning{T. Paneque-Carre\~no et al.}
    \author{T. Paneque-Carre\~no \inst{1,2},
          A. Miotello \inst{1},
          E. F. van Dishoeck \inst{2,3},
          B. Tabone \inst{4},
          A. F. Izquierdo\inst{1,2},
          S. Facchini\inst{5}
          }
          
    \institute{European Southern Observatory, Karl-Schwarzschild-Str 2, 85748 Garching, Germany
    \and Leiden Observatory, Leiden University, P.O. Box 9513, NL-2300 RA Leiden, the Netherlands
    \and Max-Planck-Institut für extraterrestrische Physik, Gießenbachstr. 1 , 85748 Garching bei München, Germany
    \and Universit\'e Paris-Saclay, CNRS, Institut d’Astrophysique Spatiale, 91405 Orsay, France
    \and Dipartimento di Fisica, Università degli Studi di Milano, Via Celoria, 16, Milano, I-20133, Italy\\
    \\
              \email{tpaneque@eso.org}}
   \date{}

 
  \abstract
   {The specific location from where molecules emit in a protoplanetary disk depends on the system properties. Therefore, directly constraining the emitting regions radially, azimuthally and vertically is key towards studying the environment of planet formation. Due to the difficulties and lack of high resolution observations, most of the current observational constraints for the vertical distribution of molecular emission rely on indirect methods.}
   {We aim to directly trace the vertical location of the emitting surface of multiple molecular tracers in protoplanetary disks. Our sample of disks includes Elias 2-27, WaOph 6 and the sources targeted by the MAPS ALMA Large Program. The set of molecules studied include CO isotopologues in various transitions, HCN, CN, H$_2$CO, HCO$^+$, C$_2$H and c-C$_3$H$_2$.}
   {The vertical emitting region is determined directly from the channel maps using the geometrical method described in \citet{Pinte_2018_method}. This method has been used in previous studies, however we implement an accurate masking of the channel emission in order to recover the vertical location of the emission surface even at large radial distances from the star and for low-SNR lines. Parametric models are used to describe the emission surfaces and characterize any structure within the vertical profile.}
   {The vertical location of the emitting layer is obtained for 10 different molecules and transitions in HD 163296. In the rest of the sample it is possible to vertically locate between 4-7 lines. Brightness temperature profiles are obtained for the entire sample and for all CO isotopologues. IM Lup, HD163296 and MWC 480 $^{12}$CO and $^{13}$CO show vertical modulations, which are characterized and found to be coincident with dust gaps and kinematical perturbations. We also present estimates of the gas pressure scale height in the disks from the MAPS sample. Compared to physical-chemical models we find good agreement with the vertical location of CO isotopologues. In HD 163296 CN and HCN trace a similar intermediate layer, which is expected from physical chemical models. For the other disks, we find that UV flux tracers and the vertical profiles of HCN and C$_2$H are lower than predicted in theoretical models. HCN and H$_2$CO show a highly structured vertical profile, possibly indicative of different formation pathways in the case of H$_2$CO.}
   {It is possible to trace the vertical locations of multiple molecular species that trace a wide variety of physical and chemical disk properties. The distribution of CO isotopologues is in agreement with theoretical predictions and the emission is found at a wide range of vertical heights $z/r = $ 0.5-0.05. The vertical location of CO may be inversely related to the stellar mass. Other molecular lines are mostly found at $z/r \leq $0.15. The vertical layering of molecules is in agreement with theoretical predictions in some systems, but not in all, therefore dedicated chemical-physical models are needed to further study and understand the diversity of the emission surfaces. }

  \maketitle
%

\section{Introduction}

Planets form in disks of dust and gas that surround young stars. In recent years the spatial resolution and sensitivity achieved by instruments such as the Atacama Large Millimeter/submillimeter Array (ALMA) has allowed us to characterize these objects in detail. A key feature that we can only now properly study is the disk vertical structure, if data at an adequate resolution are available. Protoplanetary disks have a vertical extent that depends on the system properties such as the stellar mass and radiation field, but also on the disk temperature structure and surface density distribution \citep{Bell_1997, Dalessio_1998, Aikawa_1999, van_Zadelhoff_2001_vert_models, Walsh_2010, Fogel_2011, Rosenfeld_2013}. The vertical extent of a disk is set by hydrostatic equilibrium: the balance between stellar gravity and pressure support \citep{Armitage_2015}. Tracing the vertical structure may be used to detect deviations from equilibrium that can be related to meridonial flows in the presence of planets \citep{Morbidelli_2014, Szulagyi_2022}, shadowing due to warped structures \citep{nealon_2019} or infall of material \citep{Hennebelle_2017}, among others. While larger, millimeter-sized, grains are expected to be settled in the midplane of disks, molecular and scattered light observations can be used to recover the vertical structure of the disk \citep[e.g.][]{Pinte_2018_method, Avenhaus_2018, Villenave_2020, MAPS_Law_Surf}. 

Studying the vertical location of various molecular tracers is particularly powerful for accurately mapping the disk structure due to the sensitivity of line emission to temperature, radiation and density variations. Several works have used the excitation temperature of line emission to infer the expected vertical location of a given molecule \citep[e.g.][]{van_Zadelhoff_2001_vert_models, Dartois_2003, Teague_2020_CN, MAPS_Bergner, MAPS_Ilee}. This indirect approach is promising, but it relies on physical-chemical models and assumptions about the disk conditions at the location of the emission. Direct tracing of molecular emission layers is fundamental for comparing with current models. Previous studies have focused on edge-on disks to do this \citep[e.g.][]{Podio_2020, vanthoff_2020, Villenave_2020, RuizRodriguez_2021}, however, with the development of new analysis techniques it is now possible to directly extract the vertical surfaces from moderately inclined (35-60 $^{\circ}$) sources \citep[e.g.][]{Pinte_2018_method, Paneque-Carreno_Elias1, Rich_2021, MAPS_Law_Surf, Law_2022_12CO}. Analysing systems that host moderately inclined disks allows us to study them in three dimensions, as it is also possible to trace the azimuthal and radial distribution of material.

Of the handful of moderately inclined systems with direct extraction of their vertical surfaces, the vast majority have been analyzed using only CO isotopologue emission \citep{Pinte_2018_method, Paneque-Carreno_Elias1, Rich_2021, MAPS_Law_Surf, leemker_2022_LkCa15, Law_2022_12CO}. The exceptions are AS 209, HD163296 and MWC 480 where the vertical location of HCN and C$_2$H emission was constrained in the inner $\sim$150\,au \citep{MAPS_Law_Surf}. However, the emission extends to over 300\,au \citep{MAPS_Law_radial}, therefore a large portion of the vertical structure is still unconstrained. For Elias 2-27, the vertical location of CN has also been directly measured from the channel maps and by comparing its location to that of CO isotopologues, the density and optical depth conditions of the CN emission were constrained \citep{Paneque-Carreno_2022_Elias_CN}. $^{12}$CO and $^{13}$CO emission is expected to trace optically thick slabs of material in distinct vertical regions, while C$^{18}$O likely traces closer to the midplane and CO freeze-out region \citep{van_Zadelhoff_2001_vert_models,Miotello_2014}. Through the study of CO isotopologues it is possible to trace the temperature gradient in the vertical direction \citep{MAPS_Law_Surf} and study CO abundance variations \citep{MAPS_Zhang}.

Tracing the location of less abundant molecules can offer complimentary constraints on the disk structure and the physical-chemical processes that construct the environment conditions of planet formation. Emission from CN, HCN and C$_{2}$H is expected to originate from UV-dominated regions \citep{Aikawa_2002, Cazzoletti_2018_CN, Visser_2018, MAPS_Bergner, MAPS_Guzman}. For these molecules, models predict that the emission should arise from elevated regions \citep{van_zadelhoff_2003, Cazzoletti_2018_CN, Cleeves_2018, MAPS_Bergner, MAPS_Guzman}. HCO$^{+}$ is a molecular ion, its abundance is expected to be dominant in the CO molecular layer and combined with the CO isotopologue distribution it can be used to study the disk ionization \citep{MAPS_Aikawa}. Formaldehyde (H$_{2}$CO) has two main formation pathways, gas-phase chemistry \citep{Loomis_2015} or grain-surface chemistry through the hydrogenation of CO ices \citep{Hiraoka_1994, Fuchs_2009}. H$_{2}$CO could be a better tracer of cold gas in the outer disk if it originates from desorption processes off the dust grains, however the emission may also arise from warmer and higher layers if gas-phase chemistry is dominant in its formation \citep{MAPS_Guzman, Loomis_2015}. Direct tracing of the vertical location of the H$_{2}$CO  emission could shed light on the dominant mechanism leading to the formation of H$_{2}$CO.

In this work we recover the location of the emitting regions for a sample of seven disks, using data from the MAPS ALMA Large Program \citep[\#2018.1.01055.L, ][]{MAPS_Oberg, MAPS_Czekala} and archival ALMA data of Elias 2-27 (\#2016.1.00606.S and \#2017.1.00069.S, P.I. L. P\'erez) and WaOph\,6 (\#2015.1.00168.S, P.I. G. Blake). The data selected for all the disks has sufficient signal-to-noise (SNR) and resolution (spectral and spatial) to recover the vertical structure from multiple CO isotopologues and additional molecular tracers in each system. Elias 2-27 and WaOph\,6 have spiral structures in the dust emission that have been studied at high angular resolution \citep{DSHARP_Andrews, DSHARP_Huang_Spirals} and may be originated by gravitational instabilities \citep{Perez_2016_Elias, DSHARP_Huang_Spirals, Paneque-Carreno_Elias1, Veronesi_2021_Elias}. IM Lup, GM Aur, HD 163296, MWC 480 and AS 209 are sources studied by the MAPS collaboration and each present a variety of dust and gas substructure \citep{MAPS_Oberg, MAPS_Law_radial}. Our complete sample is composed of a broad range in stellar parameters (0.5 - 2 M$_{\odot}$) and disk substructures (spirals, rings, gaps). The goal of this study is to offer observational constraints on the vertical location of the emission for a wide molecular reservoir, in a heterogeneous sample and try to relate the measured system properties to theoretical predictions.

The remainder of this paper is organized as follows. Section 2 details the data that were used and the methodology for extracting the vertical profiles. Section 3 presents our results: we first highlight the case of HD 163296, where we obtain vertical and radial profiles for ten molecules; then we show the CO emission location for all disks and study vertical modulations. This is followed by results on the radial brightness temperature profiles and an estimate of the disk gas pressure scale height. Finally, we present results for multiple molecules, other than CO, and detail our findings on the structured vertical profiles of HCN and H$_2$CO. Section 4 presents a discussion on our main results, comparing our observational constraints to theoretical estimations and in Section 5 the main conclusions of this study are highlighted.

\section{Observations and Method}

\subsection{Data}
In this work we use the publicly available data from the MAPS ALMA Large program \citep{MAPS_Oberg}. For each disk in the sample we download the image cubes that were produced using the JvM correction \citep{MAPS_Czekala}. An initial visual assessment is done to determine if a specific molecule/disk can be used for extracting the vertical structure from the channel maps, following the methodology detailed in section 2.2. Reasons for rejecting a molecule can be low SNR or lack of resolution to identify a Keplerian morphology in the channel emission. Table \ref{MAPS_files} shows the details of the selected files for each tracer , corresponding to those molecules for which it is possible to conduct our analysis in at least three disks. The data set used may be identified by either the robust parameter or the beam size. For each molecule, the selected robust or beam size is the same for all disks in the MAPS sample. For datasets selected based on the robust parameter the difference in beam value between major and minor axis is typically of order $\leq$0.02\arcsec. Between sources there is also a slight beam size difference of $\leq$0.02\arcsec. At a distance of $\sim$120\,au this represents a difference of $\sim$2.5\,au, therefore we consider beam size variations negligible in our calculations of the vertical profile. In addition to the information presented in Table \ref{MAPS_files}, for HD\,163296 we also study CN ($N =1-0 $, 113.499\,GHz) and c-C$_{3}$H$_{2}$ ($J =7_{07}-6_{16} $, 251.314\,GHz). The selection of these data is done by the beam size, 0.5\arcsec for CN and 0.3\arcsec for c-C$_{3}$H$_{2}$. A detailed study on HD\,163296 is presented in section 3.1. 

\begin{table}[h]
    \centering
    \def\arraystretch{1.3}
    \setlength{\tabcolsep}{4pt}
    \caption{Selected files for molecules in MAPS sample disks}
    \begin{tabular}{c| c |c|c|c}
    \hline
    \hline
         Molecule & Transition & Freq. [GHz] & robust & beam $^a$ [\arcsec]\\ 
    \hline
    $^{12}$CO & $J=2-1$ &  230.538 & 0.5& $\sim$0.13\\
    $^{13}$CO & $J=2-1$ & 220.398 & 0.5& $\sim$0.13\\
    C$^{18}$O & $J=2-1$ & 219.560 & 0.5& $\sim$0.13\\
    $^{13}$CO & $J=1-0$ & 110.201 & 0.5& $\sim$0.26\\
    \hline
    HCN & $J=3-2$  & 265.886 & - & 0.30\\
    C$_{2}$H & $N=3-2$ & 262.040& - &0.15\\
    H$_{2}$CO & $J=3_{03}-2_{02}$ & 218.222 & - & 0.30 \\
    HCO$^{+}$ & $J=1-0$ & 89.188 & 0.5 & $\sim$0.32\\
    
    \hline 
    \end{tabular}
    \tablefoot{
    \tablefoottext{a}{For molecules that were selected based on the robust parameter, the beam value corresponds to the mean value of all sources. In all $J = 2-1$ transitions the typical deviation from the mean value between sources is 6.7\% and in the $J = 1-0$ transitions it is 8.3\%}
    }
   
    \label{MAPS_files}
\end{table}

\begin{table*}[h!]
    \centering
    \def\arraystretch{1.3}
    \setlength{\tabcolsep}{6pt}
    \caption{Studied systems, their properties and the analyzed molecules in each case.}
    \begin{tabular}{c|c|c|c|c|c|c}
    \hline
    \hline
         Star $^a$& M$_*$ & L$_*$ & Distance & Inclination & PA & Studied Molecules $^b$ \\ 
          & [M$_{\odot}$] & [L$_{\odot}$] & [pc] & [deg] & [deg] &  \\ 
    \hline
    IM Lup & 1.1 &  2.57 & 158& 47.5 & 144.5 & $^{12}$CO, $^{13}$CO, C$^{18}$O, HCN, HCO$^+$, H$_2$CO \\
    GM Aur & 1.1 & 1.2 & 159&  53.2 & 57.2 & $^{12}$CO, $^{13}$CO, C$^{18}$O, HCN, HCO$^+$, H$_2$CO \\
    AS 209 & 1.2 & 1.41 & 121 & 35.0 & 85.8 & $^{12}$CO, $^{13}$CO, C$^{18}$O, HCN, HCO$^+$, H$_2$CO, C$_2$H \\
    HD 163296 & 2.0 & 17.0 & 101& 46.7 & 133.3 &  $^{12}$CO, $^{13}$CO, C$^{18}$O, HCN, CN, HCO$^+$, H$_2$CO, C$_2$H, c-C$_3$H$_2$  \\
    MWC 480 & 2.1 & 21.9 & 162 & 37.0& 148.0 & $^{12}$CO, $^{13}$CO, C$^{18}$O, HCN, HCO$^+$, H$_2$CO, C$_2$H \\
    \hline
    Elias 2-27 & 0.46  & 0.91 & 116 & 56.7 & 118.8 & $^{12}$CO ($2-1$), $^{13}$CO ($3-2$), C$^{18}$O ($3-2$), CN ($7/2 - 5/2$) \\
    WaOph 6 & 1.1 & 0.68 & 123 &47.3 & 174.2 & $^{12}$CO ($3-2$), $^{12}$CO ($2-1$), $^{13}$CO ($3-2$),  HCO$^+$ ($3-2$)\\

    \hline 
    \end{tabular}
    \tablefoot{
    \tablefoottext{a}{Stellar and disk parameters taken from \citet{MAPS_Oberg} for the 5 disks in the MAPS sample and from \citet{DSHARP_Huang_radial} and \citet{DSHARP_Andrews} for Elias\,2-27 and WaOph\,6, except for stellar masses. The stellar mass is estimated dynamically in \citet{Veronesi_2021_Elias} for Elias 2-27 and in \citet{Law_2022_12CO} for WaOph\,6.}
    \tablefoottext{b}{The transitions of each molecule for the disks of the MAPS sample are detailed in Section 2.1. For Elias 2-27 and WaOph\,6 the $J$ transition of each molecule is shown in parenthesis.}
    }

    \label{table_sample_all}
\end{table*}

To the MAPS sample, we add publicly available data of Elias 2-27 and WaOph 6. In both sources we study $^{12}$CO $J=2-1$ emission from DSHARP \citep{DSHARP_Andrews}. We also include $^{13}$CO and C$^{18}$O $J=3-2$ data of Elias 2-27 from ALMA programs \#2016.1.00606.S and \#2017.1.00069.S (P.I. L. P\'erez). Calibration and imaging procedures of Elias 2-27 can be found in \citet{DSHARP_Andrews} and \citet{Paneque-Carreno_Elias1}. For WaOph 6 we include archival data of ALMA program \#2015.1.00168.S (P.I. G. Blake) and the ALMA Large Program DSHARP \citep{DSHARP_Andrews}. After self-calibration we can detect emission from $^{12}$CO, $^{13}$CO, C$^{18}$O, HCN, CN and HCO$^{+}$ $J = 3-2$ in WaOph 6. Due to the moderate spatial resolution of the data set (0.3-0.4$\arcsec$), in this study we will only analyze $^{12}$CO, $^{13}$CO and HCO$^{+}$ data, where we can confidently study the vertical distribution of the disk. The integrated moment maps for all molecules observed in WaOph 6 and details on the self-calibration steps are found Appendix A.

Overall, our sample consists on two Herbig and four T-Tauri stars \citep{MAPS_Oberg, DSHARP_Andrews}. There are three disks with spirals, and four with rings and gaps \citep{DSHARP_Huang_radial, DSHARP_Huang_Spirals}. Additionally, various sources show distinct kinematical features such as signatures of late infall in GM Aur \citep{MAPS_huang} and Elias 2-27 \citep{Paneque-Carreno_Elias1} and possible planetary perturbations in HD 163296 \citep{Pinte_2018_hd16planet, Teague_2018_hd16planet, Izquierdo_2021_hd16planet, MAPS_Teague} and MWC 480 \citep{MAPS_Teague}.

\subsection{Method}
\subsubsection{Vertical profile extraction}
To extract the emission surfaces we use the geometrical method outlined in \citet{Pinte_2018_method} as applied in \citet{Paneque-Carreno_Elias1}. Knowing the central star position and geometrical properties, the vertical profile of the upper emitting layer can be recovered directly from the channel map observations by tracing the location of the maxima of emission along each channel, assuming that they are tracing the isovelocity curve. For the MAPS sample, the data have been centered and aligned following the location of the continuum peak emission \citep{MAPS_Oberg}, where we assume the central star is located. This same centering has been performed on the Elias 2-27 and WaOph-6 datasets, therefore we consider the center of the image as the stellar location. If the inclination, flaring and resolution of the disk allow the identification of upper and lower surfaces it is possible to confidently trace independently both vertical profiles. In cases when the surfaces can not be visually identified separately we assume that the peak of emission of the channel map comes from the upper surface, which is expected to be brighter than the lower surface. 

\begin{figure}[h!]
   \centering
   \includegraphics[width=\hsize]{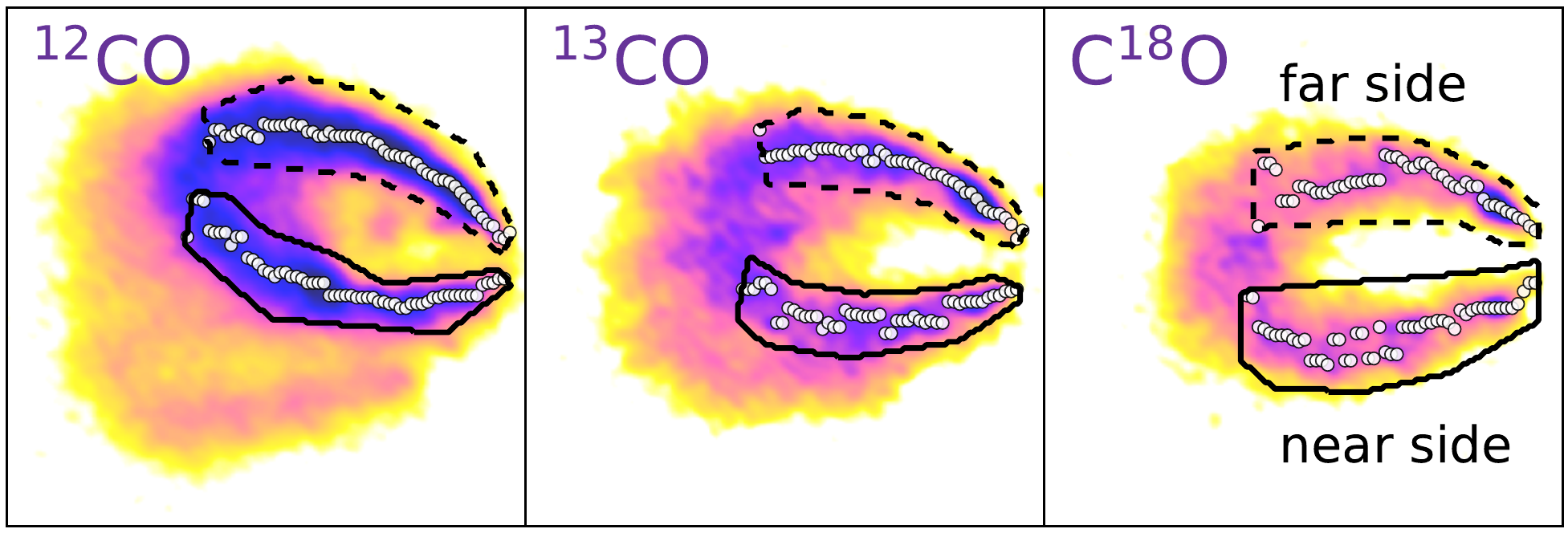}
      \caption{Rotated 7.86\,km\,s$^{-1}$ channel map emission of the $J = 2-1$ transition of CO isotopologues in HD 163296. Overlayed are the regions masked as far and near sides of the upper surface in dashed and solid black lines, respectively. White dots trace the emission maxima within the masked regions.
              }
         \label{masks}
\end{figure}

From the extracted location of the maxima, geometrical relations can be used to obtain the vertical profile of the emission \citep[see][for more details]{Pinte_2018_method}. To trace the maxima we use our own implementation of the method (ALFAHOR), which relies on masking each channel through visual inspection to identify separately the near and far sides of the emission from the upper layer (see Figure \ref{masks} for clarification). The masks are selected after the channel map emission has been rotated, using the position angles indicated in Table \ref{table_sample_all}, such that the bottom side of the disk emission is towards the South.  The maxima of emission are retrieved automatically, but only from sampling the pixels within the masks. The masks are selected with as conservative margins as possible to avoid a bias in the recovered structure and there are no particular criteria regarding maximum outer radius or any other value\footnote{a repository with the mask and code developed for this study can be found in https://github.com/teresapaz/alfahor.}. In this work, we only study the upper layer of emission, but in those cases where the lower layer can be identified (as can be observed for $^{12}$CO in Figure \ref{masks}) it would be possible to additionally trace the vertical extent of the lower layer.

Previous work for some of the disks and molecules in our sample \citep{MAPS_Law_Surf} used a publicly available implementation of the \citet{Pinte_2018_method} method, DISKSURF \citep{disksurf}. In DISKSURF the search of the emission maxima can be blind or constrained with initial expected values of the minimum and maximum $z/r$, the minimum SNR and selected channels. While this allows for a systematic search and characterization of the channel emission, in some cases pixels from the bottom surface are incorrectly classified as located in the upper surface, and the other way around. This induces noisier vertical profiles, as there is a large data spread caused by the contamination of pixels from the lower surface of the disk. The SNR cut-off is also problematic when dealing with emission from less abundant molecules. By masking the channels after visual inspection, we reach locations of lower SNR and avoid contamination from the lower surface, obtaining a more accurate description of the vertical profile. Our implementation has been tested and illustrated in previous works \citep{Paneque-Carreno_Elias1, leemker_2022_LkCa15} and in this analysis we also compare it to the results of \citet{MAPS_Law_Surf} using DISKSURF (See Appendix B). Our results are consistent for bright tracers such as $^{12}$CO and $^{13}$CO, particularly in the inner region ($<$200\,au), however, through our methodology we are able to trace a larger inventory of molecules to further radii.

For each disk and molecule we obtain the maxima from the channel maps by sampling every quarter of the beam semi-major axis in cartesian coordinates, after correcting for the disk position angle (PA). From all the obtained data points (see Appendix B panels with all of the retrieved data points for each disk and tracer) we present the vertical profiles and the associated dispersion as the average value and the standard deviation within radial bins as wide as the beam semi-minor axis. In some disks, due to low SNR, there are less data points. To avoid biases due to lack of sampling, we only consider the data from radial bins where there are at least two independent data points. For Elias 2-27 it has been shown that there are azimuthal asymmetries in the vertical extent \citep{Paneque-Carreno_Elias1}, but we do not find any indication of this behaviour in any of the other disks and molecules of our sample. Therefore all of the data points are considered to compute the radial profiles. In the case of Elias 2-27 only the data points from the west side are considered. Table \ref{table_sample_all} details for each system the molecular emission from which we are able to produce vertical profiles.

\begin{figure*}[h!]
   \centering
   \includegraphics[width=\hsize]{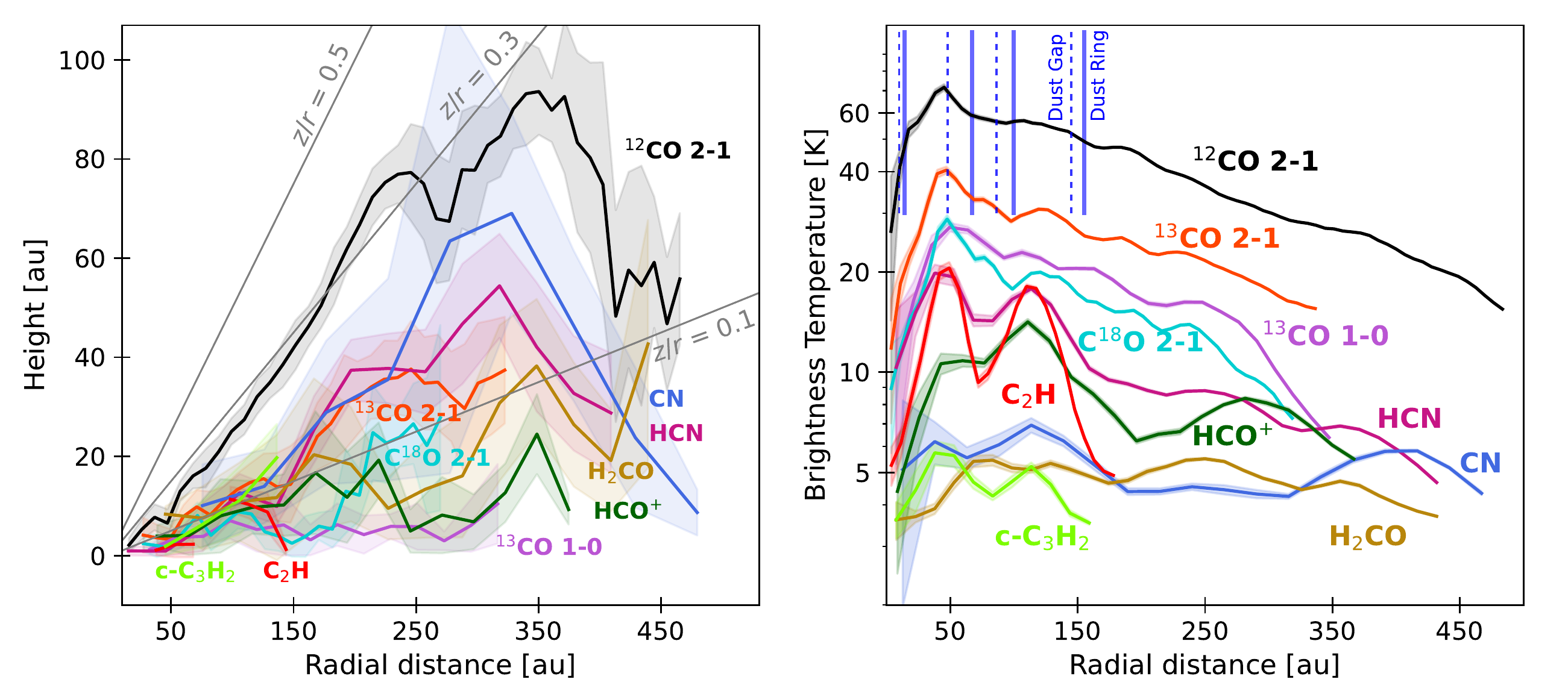}
      \caption{For HD 163296, vertical emission profiles in the left panel and azimuthally averaged peak brightness temperature profiles in the right panel for various tracers, beyond CO isotopologues. Solid colored lines show the mean value in each profile and shaded regions the 1$\sigma$ data dispersion. The vertical blue lines in the right panel indicate the location of millimeter continuum gaps and ring detected by \citet{DSHARP_Huang_radial}. 
              }
         \label{panel_hd16}
\end{figure*}

Each emission surface in each disk and isotopologue can be parametrized using an exponentially tapered power law defined as,

\begin{equation}
z(r) = z_0\times \left(\frac{r}{100\,\mathrm{au}}\right)^{\phi} \times \exp\left[\left(\frac{-r}{r_{\mathrm{taper}}}\right)^{\psi}\right]
\end{equation}

The best fit values for $z_0$, $\phi$, $r_{\mathrm{taper}}$ and $\psi$ in each system are found by using a Monte Carlo Markov Chain (MCMC) sampler as implemented by emcee \citep{emcee_ref}. For the fitting procedure, we consider all data points. If convergence is not reached for $r_{\mathrm{taper}}$ and $\psi$, or if $r_{\mathrm{taper}}$ is much larger than the radial extent of the profiles, we assume a simple power-law profile and fit only for $z_0$ and $\phi$. Table \ref{table_vertical_co} presents the computed parameters of the exponentially tapered or single power laws for each disk and CO isotopologue.

\subsubsection{Brightness temperature calculation}

The brightness temperature ($T_b$) profiles are obtained from the peak intensity ($I$) using the complete Planck law. The relationship between both parameters is the following,

\begin{equation}
    T_b = \frac{h \nu}{k} \ln^{-1}\left(1 + \frac{2h\nu^3}{I c^2}\right)
\end{equation}

where $h$ is the Planck constant, $k$ the Boltzmann constant, $c$ the speed of light and $\nu$ the frequency of the emission. The temperature profiles are computed from the azimuthally averaged peak intensity maps. To accurately deproject the data, we use the best-fit power-law or exponentially tapered power-law model of each molecule. The azimuthally averaged peak intensity profile is obtained using the GoFish package \citep{gofish}, a $\pm$30$^{\circ}$ wedge across the semi major axis \citep[as done in][]{MAPS_Law_radial} and radial bins half the size of the beam semi-major axis. 

\section{Results}

\subsection{The special case of HD\,163296}

HD\,163296 stands out from the rest of the sample, as it is the disk where we were able to trace the emission layer for the highest number of tracers (see Table \ref{table_sample_all}). For instance, it is the only system from the MAPS sample where we could obtain vertical profiles for CN and c-C$_3$H$_2$. HD\,163296 is also a system of interest due to the detection of at least two planetary signatures at 94 and 261\,au \citep{Pinte_2018_hd16planet, Teague_2018_hd16planet, MAPS_Teague, Izquierdo_2021_hd16planet}.  

Figure \ref{panel_hd16} presents the emission surface and brightness temperature profiles of each studied molecule. The emission surface of $^{12}$CO has the highest $z/r$ and traces a value of 0.3 up to $\sim$\,350\,au where it has a turning point and steeply declines. The emission surfaces of CN, HCN and $^{13}$CO $J=$2-1 are located in the interval of $z/r \sim$ 0.1-0.3. c-C$_3$H$_2$ and H$_2$CO trace close to $z/r \sim$ 0.1. C$_2$H also follows $z/r \sim$ 0.1, but its emission surface rapidly declines at $r\sim$100\,au. This is in agreement with the C$_2$H vertical profile derived in \citet{MAPS_Law_Surf} for HD\,163296. Below $z/r \sim$ 0.1 is the emission of HCO$^+$, $^{13}$CO $J=$1-0 and C$^{18}$O (only for the inner $\sim$150\,au in the case of C$^{18}$O). Beyond $\sim$200\,au, the vertical profile of C$^{18}$O elevates and traces a region closer to $^{13}$CO $J=$2-1

The right panel of Figure \ref{panel_hd16} shows our results for the brightness temperature profiles of each tracer. By comparing molecules found at similar vertical locations, but with different brightness temperatures, we can estimate the optical depth of the emission \citep[not presented in this work, but see][for an example]{Paneque-Carreno_2022_Elias_CN}. In the case of optically thick tracers, the brightness temperature will be a probe of the kinetic temperature of the studied region.

\begin{figure*}[h!]
   \centering
   \includegraphics[width=\hsize]{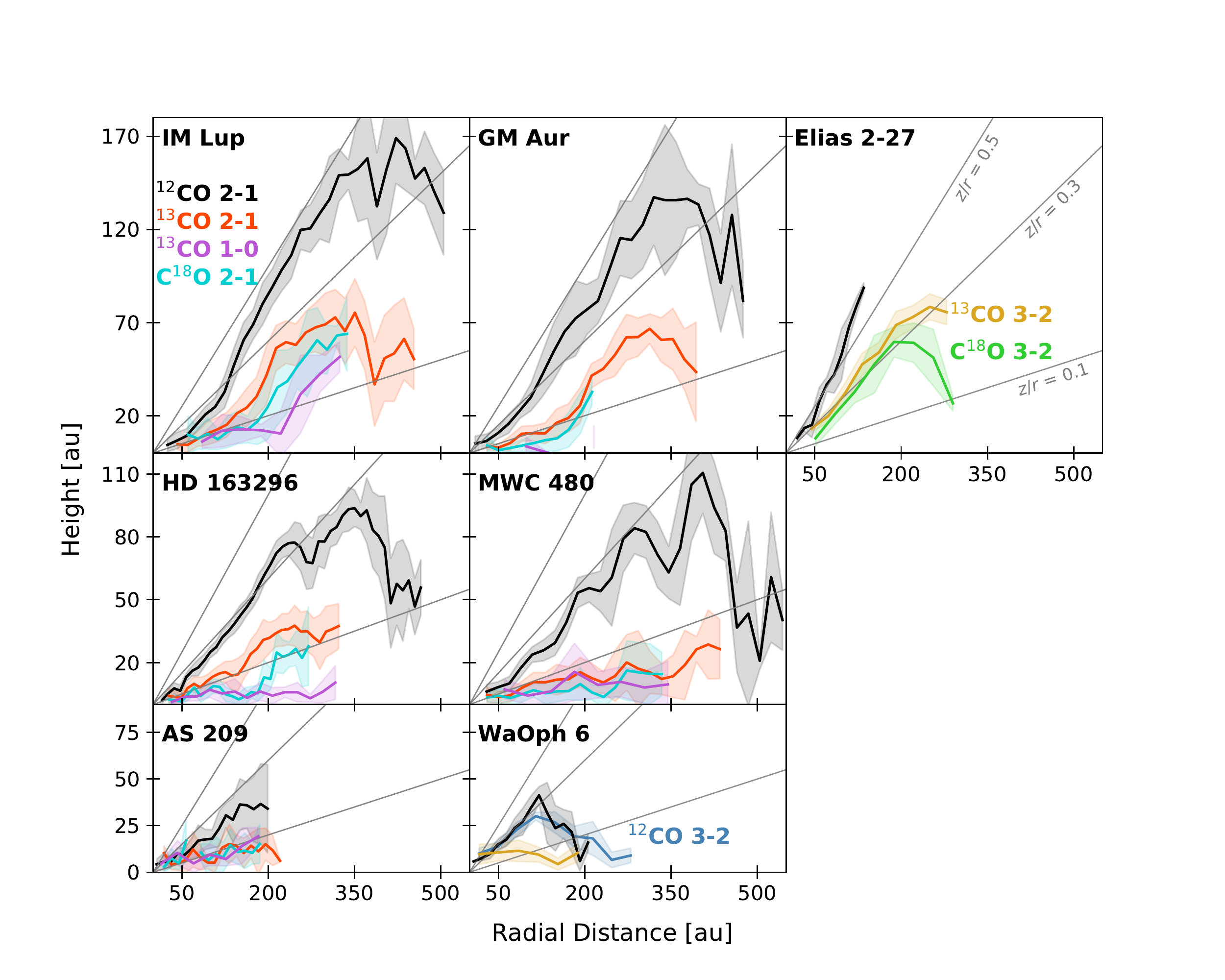}
      \caption{Vertical profiles for CO isotopologues as extracted from the channel maps of each disk. Shaded region shows the dispersion of the data points and solid colored line traces the average values within each radial bin. Note that Elias 2-27 and WaOph 6 were observed at a higher transition ($J = 3-2$) in some CO isotopologues. Solid grey lines show constant $z/r$ of 0.1, 0.3 and 0.5. Each row has a different vertical extent.
              }
         \label{panel_CO_all}
\end{figure*}

As expected, due to its location, high abundance and optical depth, $^{12}$CO has the highest brightness temperature. CN, HCN and $^{13}$CO $J=2-1$ emit from the same vertical region, however, $^{13}$CO $J=2-1$ has a temperature of $\sim$20\,K, HCN is found at $\sim$10\,K and CN at $\sim$5\,K. We expect $^{13}$CO $J=2-1$ to be marginally optically thick, while theoretical models and observational constraints \citep{Cazzoletti_2018_CN, MAPS_Bergner} predict that CN and HCN are likely optically thin, which would lead to a lower brightness temperature, as measured in this work. For the $z/r \sim$ 0.1 molecules, we trace similar temperatures for c-C$_3$H$_2$ and H$_2$CO at $T_b\sim$5\,K and a higher temperature for C$_2$H, with a strong dip at 75-85\,au. This dip is also seen in the HCN and c-C$_3$H$_2$ profiles and coincides with a gap found in the line emission of these molecules \citep{MAPS_Law_radial}. Finally in the molecules closer to the midplane, $^{13}$CO $J=1-0$ and C$^{18}$O have a similar $T_b\sim$18\,K. HCO$^+$ has a lower temperature profile, indicating it is likely optically thin with $T_b\sim$8\,K.

Besides optical depth differences, \citet{leemker_2022_LkCa15} show that differences in the spectral resolution may lead to variations in the extracted brightness temperature. Lower spectral resolution data will result in lower brightness temperature ranging from 10-60\% depending on the resolution difference \citep[see Appendix A.3 in][]{leemker_2022_LkCa15}. From the MAPS data, all observations of Band 6 (211-275\,GHz) have a velocity resolution of 0.2\,km\,s$^{-1}$ whereas observations taken in Band 3 (84-116\,GHz) have a resolution of 0.5\,km\,s$^{-1}$. Therefore, CN, HCO$^{+}$, and $^{13}$CO $J=1-0$ would likely have higher brightness temperature profiles if we compared them at the same spectral resolution as the rest of the tracers. The brightness temperature of all molecules shows a steep turnover in the inner 50\,au, which is likely due to effects of beam smearing. At the source distance, the beam major axis of each tracer takes a value between 13-30\,au and 50\,au in the case of CN. Additional perturbations in the temperature profiles may be due to line flux subtraction from the continuum emission, as we are using the continuum-subtracted datasets. The dust features are indicated in the top of the right panel of Fig. \ref{panel_hd16} for reference.

\subsection{CO isotopologue vertical structure of the full sample}

The extracted emission surfaces of the CO isotopologues from the complete sample of disks are presented in Figure \ref{panel_CO_all}. As explained in section 2.2, we are able to trace the emitting layer out to a larger radial extent than the previous work on the MAPS sample \citep{MAPS_Law_Surf}. The main difference is in the C$^{18}$O $J =$ 2-1 emission, which at larger radii is observed to trace a region very similar to $^{13}$CO $J = 2-1$. This is seen in all disks where both isotopologues are available and is in agreement with previous IM Lup  \citep{Pinte_2018_method} and Elias 2-27 \citep{Paneque-Carreno_Elias1} analysis. We are also able to compute a vertical profile for $^{13}$CO $J =$ 1-0, which mostly traces a layer below $^{13}$CO $J =$ 2-1, except in MWC 480 and AS 209 where both transitions are very close to the midplane.

The sample is divided in three groups, indicated by the three separate rows of Figure \ref{panel_CO_all}. This classification depends on the $z/r$ values that the $^{12}$CO traces and the radial extension of the emission. IM Lup, GM Aur and Elias 2-27 are the most vertically extended disks, with $z/r \geq$0.3. HD\,163296 and MWC\,480, the two Herbig Ae disks, have a $^{12}$CO vertical profile that traces $z/r \sim$0.3. Finally AS\,209 and WaOph\,6 are the flattest and least radially extended disks in our sample, with $z/r \sim$ 0.1-0.3. 

In some cases, such as for $^{12}$CO in Elias 2-27, we lack sampling of the outer regions due to cloud absorption \citep{Perez_2016_Elias, Paneque-Carreno_Elias1}. In this case the obtained best-fit for the vertical profile (methodology described in section 2.2) follows a single power-law. However this description of the data may only be valid in the inner $\sim$100\,au of the disk. For C$^{18}$O and $^{13}$CO ($J =$ 1-0) it is necessary to fit with a single power-law model in all disks due to the flat morphology or lack of turnover in the measured profile. It is important to emphasize that this does not mean there is no turnover as our parametric description is only valid up to the sampled radial location. 

\subsubsection{Modulations in the surface and correlation to kinematical and dust features.}

\begin{figure}[h!]
   \centering
   \includegraphics[width=\hsize]{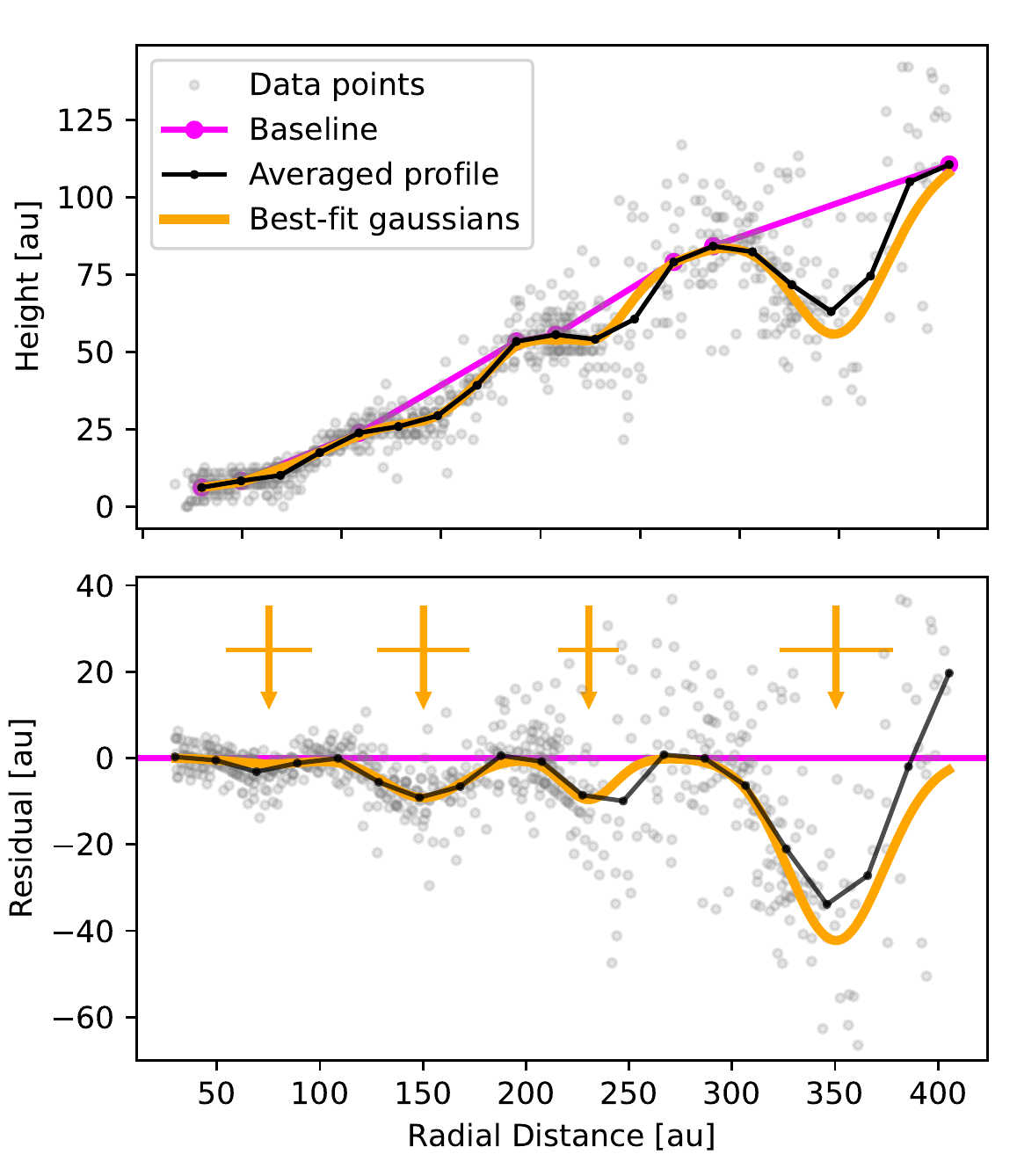}
      \caption{Example of baseline and best-fit gaussian modulation extraction for the MWC\,480 $^{12}$CO data set
              }
         \label{example_mod}
\end{figure}

Even though the data can be fitted using a power-law, the vertical profiles in Figure \ref{panel_CO_all} show modulations or "bumps". This behaviour is identified in the $^{12}$CO surface of HD\,163296, MWC\,480 and IM\,Lup. To trace the deviations from the possibly smooth vertical profile a reference profile is assumed, which we refer to as ``baseline'' (see Fig. \ref{example_mod}). This baseline does not coincide with the previously derived best-fit power-law or exponential profiles, which are not considered for this analysis because they cut through some of the features seen in Figure \ref{panel_CO_all}. It is important to note that the retrieval of vertical modulations will strongly depend on the assumed baseline, which we cannot know with certainty without dedicated modeling and understanding of the properties of each disk. 

We assume a baseline that traces the averaged vertical profile such that it follows a positive gradient at all radial locations. If there is a turnover, after which modulations are observed, we allow the baseline to follow only negative gradients, from the location where no further positive gradients are found. The final baseline profile is substracted from all the data points and we compute the binned residuals. It is assumed that data with lower vertical values with respect to the baseline trace modulations and we fit gaussians at these locations. The amount of gaussian features to be fitted and the initial guess on the radial location are determined through visual inspection of the residual data points. The best-fit surface is obtained by simultaneously fitting multiple gaussians to the assumed baseline. The fit is done considering all of the retrieved data points from the channel map analysis, the binned data are only used to determine the baseline and to aid visual inspection. This process is schematically shown in Figure \ref{example_mod}.

\begin{figure*}[h!]
   \centering
   \includegraphics[width=\hsize]{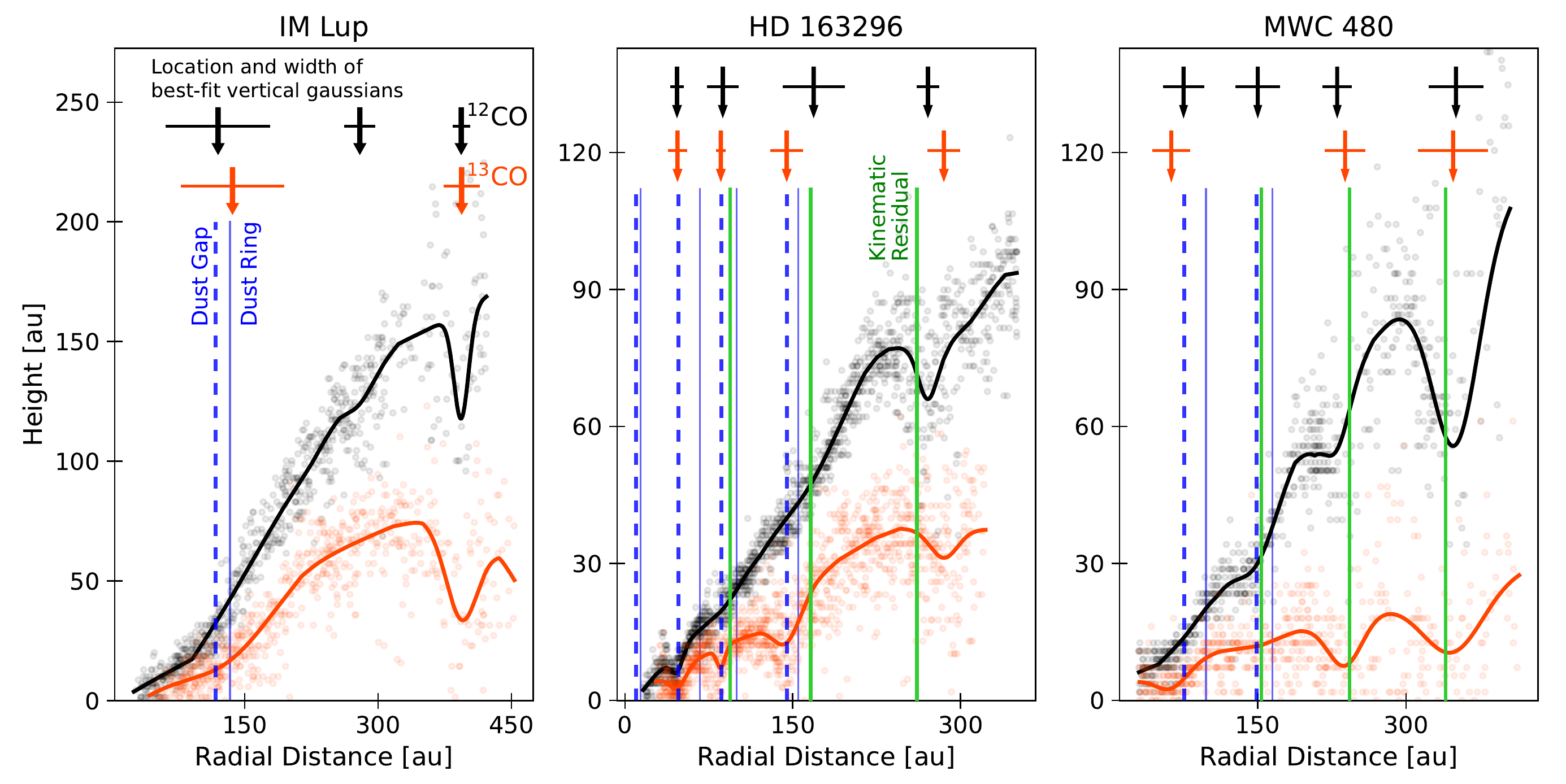}
      \caption{Data corresponding to $^{12}$CO $J = 2-1$ in black and $^{13}$CO $J = 2-1$ in orange for three disks where modulations in the vertical surface are detected. Colored dots indicate the extracted data from the channel maps for each tracer. Continuous lines trace the best-fit surface considering the reference baseline and a number of fitted gaussians. The location and width of the gaussians for each molecule are signaled at the top of each panel. Vertical blue lines trace the dust structure from \citet{DSHARP_Huang_radial}. Vertical green lines show the reported kinematic residuals from \citet{MAPS_Teague} and \citet{Izquierdo_2021_hd16planet}.
              }
         \label{CO_bumps}
\end{figure*}

The $^{12}$CO and $^{13}$CO $J = 2-1$ isotopologues are analyzed for HD\,163296, MWC\,480 and IM\,Lup, which are the disks and tracers where this behaviour is most clearly identified. C$^{18}$O is also analysed for HD\,163296 and MWC\,480, but as no modulations are found in IM Lup the analysis is not shown in this section. Detailed plots and values for each tracer and disk can be found in Appendix B. The best-fit models of $^{12}$CO and $^{13}$CO are shown overlayed to the data in Figure \ref{CO_bumps}, where we compare the locations of the modulations in different tracers with the location of the dust rings and gaps found in ALMA millimeter continuum emission images \citep{DSHARP_Huang_radial} and reported kinematic features \citep{MAPS_Teague, Izquierdo_2021_hd16planet}. 

HD\,163296 displays a tight correlation between all of the modulations in both tracers. Additional analysis of the C$^{18}$O emission in HD\,163296 also finds modulations at similar radial locations as the inner three structures reported in $^{12}$CO and $^{13}$CO. In MWC\,480 we find a feature in $^{12}$CO and C$^{18}$O that does not seem to have a correspondent in $^{13}$CO, at 150\,au. IM\,Lup has a strong feature at 393\,au that is recovered in both tracers, however the other modulations do not relate between isotopologues. Contrary to the other disks, IM\,Lup does not have visually identifiable drops in the vertical height in the inner 300\,au and the features we obtain are broader than in the other systems. It may be that in this case we are tracing the flaring of the inner disk that due to our linear baseline is not considered properly in our reference surface model.

It is found that HD\,163296 and MWC 480 present a strong correlation between the location of millimeter continuum emission gaps \citep{DSHARP_Huang_radial} and vertical modulation in the gas emission. A correlation is also found between the radial location of kinematic deviations detected in residuals from the velocity maps \citep{MAPS_Teague, Izquierdo_2021_hd16planet} and the vertical modulations traced in both HD\,163296 and MWC 480.  In contrast, there is no relation between the emission gaps and rings traced from the integrated emission maps \citep{MAPS_Law_radial} in either CO isotopologue for any of the disks. An additional caveat to these results may be that the extraction of the emission surfaces assumes a smooth keplerian motion of the material \citep{Pinte_2018_method} and strong deviations could be responsible for the features we detect, precisely quantifying this effect will be studied in a future work. However, the vertical modulations are recovered for the complete azimuthal range, they are not localized in azimuth as is the case for planetary kinks \citep[e.g.][]{Pinte_2018_method, Perez_2018_dopplerflip, izquierdo_2021_discminer1}. Considering that the emission surfaces extracted for $^{12}$CO and $^{13}$CO are expected to be related to the surfaces of $\tau = $1, it is likely that the recovered modulations trace actual physical perturbations and decreases in the emitting surface or column density of CO. We will comment this further in section 4.

Previous analysis of vertical modulations by \citet{MAPS_Law_Surf} studied the inner $\sim$100\,au for the MAPS sample and our results recover all of the features previously reported in HD\,163296 and MWC\,480. The depths and widths of our features slightly differ in some cases from the previous characterization due to the differences in extracting the vertical profile and in computing the assumed surface. The depth of each modulations is computed in the same way as in \citet{MAPS_Law_Surf}. We consider the height value of the assumed surface at the feature location ($z(r_0)$) and the height depletion caused by the gaussian feature ($\Delta z$), which is equivalent to the fitted amplitude. The ratio between these numbers results in a percentual description of the depletion from the assumed surface value. Higher depth values indicate a larger relative decrease in the scale height value. Our results show that, for features that are found in both tracers at similar locations, the relative depth in $^{13}$CO is larger than $^{12}$CO for all cases. Exact values are found in Appendix B.

\subsubsection{Brightness temperature profiles}

\begin{figure*}[h!]
   \centering
   \includegraphics[scale=0.72]{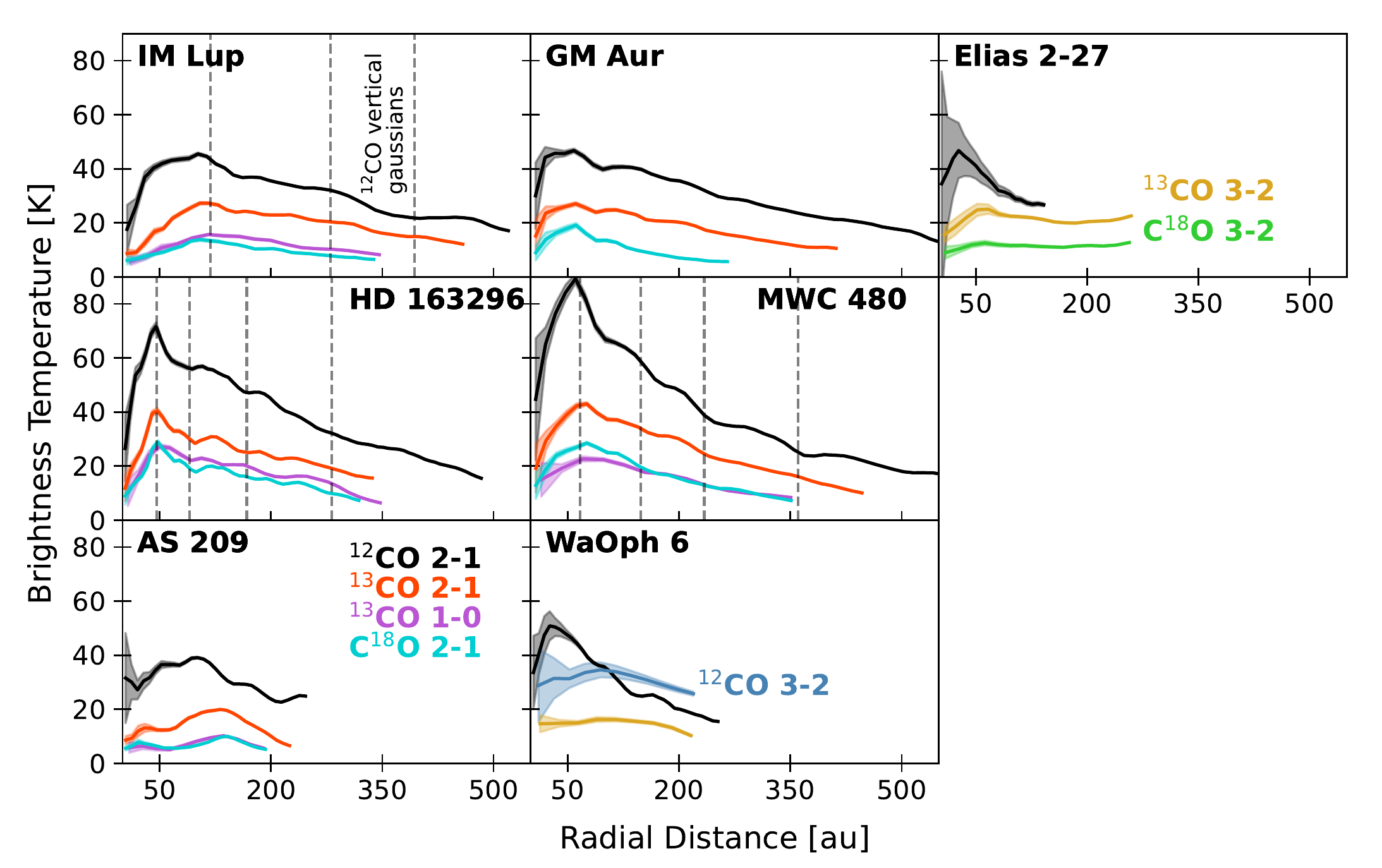}
      \caption{Brightness temperature profiles of the CO isotopologue emission in each disk. Solid line indicates the mean value and the shaded region the standard deviation within each radial bin, divided by the number of beams in each annuli. Vertical and radial scales are shared in each panel. Dashed vertical lines in IM Lup, HD 163296 and MWC 480 indicate the location of the vertical modulations determined in this work for $^{12}$CO.
              }
         \label{panel_temp}
\end{figure*}

As done in HD\,163296, we obtain the brightness temperature profiles for the CO isotopologue emission of the complete sample. In particular, for Elias 2-27 we use only the emission from the east side, due to the elevation and brightness asymmetries \citep{Perez_2016_Elias, Paneque-Carreno_Elias1}. WaOph\,6 and AS\,209 also have strong absorption and a brightness asymmetry, therefore for the azimuthal profiles we only consider the south and west sides, respectively. The rest of the systems have their profiles extracted using a $\pm$30$^\circ$ wedge across the semi major axis. 

Figure \ref{panel_temp} shows the results for each system and CO isotopologue. The ordering in the panels matches Figure \ref{panel_CO_all}. The most vertically extended disks are in the top row, the intermediate disks in the middle row and the smaller, flatter disks in the bottom. All disks show a slowly decreasing temperature profile towards larger radii. Elias 2-27, HD\,163296, MWC\,480 and WaOph\,6 show a steeper slope in the temperature decrease of $^{12}$CO. HD\,163296 and  MWC\,480 also display higher temperatures for all isotopologues, which is expected as they are both warm Herbig disks orbiting more luminous stars. In all disks from the MAPS sample $^{13}$CO $J = 1-0$ and C$^{18}$O $J = 2-1$ trace the same brightness temperatures, which are mostly below the CO freeze-out value ($\sim$ 21K). This happens particularly at radii beyond 80\,au for the Herbig stars, but at all radii for IM Lup, GM Aur and AS 209. We note that, as discussed previously, the spectral resolution of $^{13}$CO $J = 1-0$ is lower than that of the other transitions, which may artificially result in lower brightness temperatures \citep{leemker_2022_LkCa15}. Very low brightness temperatures may also be indicative of low optical depths of the studied isotopologues and do not represent the kinetic temperature of the gas at that location. High optical depth tracers, such as $^{12}$CO are expected to have  brightness temperature profiles that are better tracers of the kinetic temperature at their location.

The vertical dashed lines in Figure \ref{panel_temp} show the location of the modulations in the vertical surface of $^{12}$CO for IM Lup, HD 163296 and MWC 480 (see section 3.2.1). While the temperature profiles do not show such strong features, slight modulations are distinguished that may be coherent with the locations of the vertical modulations. We tentatively detect relations between dips in the brightness temperature profiles of $^{12}$CO and the location of the vertical modulations. Future detailed modelling of each source must be conducted to determine if the variations in the temperature profile are sufficient to explain the drops in the CO emission layer and determine the causes. 

Even though the measured $^{12}$CO vertical profiles of each disk are different and allow the classification of the sample in three categories based on vertical extension, the brightness temperature profiles of $^{12}$CO are very similar. Considering only the disks from the MAPS sample, there is a clear difference in the $^{12}$CO temperatures of T Tauri and Herbig Ae stars, as has already been reported in \citet{MAPS_Law_Surf}. However, within the T Tauri stars, the vertical extent of the $^{12}$CO emission in the disks is twice as large in IM Lup and GM Aur than in AS 209. WaOph\,6 and Elias 2-27 are also T Tauri stars and both show a warmer $^{12}$CO inner disk than the other low mass stars from the MAPS sample.

\subsubsection{Calculation of disk scale height}

We have presented the vertical location of several CO isotopologues in a sample of disks, however this emission surface depends on several physical-chemical conditions. A more transverse characterization of a disk's vertical structure is given by the gas pressure scale height ($H$), which can be used to compare theoretical models with observations. In order to derive this information, we design a simple single layer model.  Despite the simplicity of the model, this allows us to provide first quantitative estimate of the material distribution. Assuming that the disk is vertically isothermal it is possible to relate the scale height to the total volumetric gas density ($\rho_\mathrm{gas}$) and the surface density of the disk ($\Sigma$) through,

\begin{equation}
    \rho_\mathrm{gas}(z) = \Sigma \frac{e^{-z^2/2H^2}}{\sqrt{2\pi}H}
\end{equation}

We expect our derived vertical profile of $^{12}$CO to trace the region where the emission becomes optically thick ($\tau\geq1$) or where CO becomes self-shielding. For $^{12}$CO ($J = 2-1$) at a temperature of 40-60K this occurs when the vertically integrated column density of $^{12}$CO reaches the critical value $N_{\mathrm{CO}, \mathrm{crit}} \simeq 5 \times 10^{16} \mathrm{cm}^{-2}$ \citep[resulting in $\tau_{CO}\sim$1, values obtained using RADEX online simulator, ][]{van_der_Tak_2007_RADEX}. As we are studying CO column density, we assume a constant $x$(CO) abundance of 2.8$\times$10$^{-4}$ with respect to H$_2$ in the disk \citep{Lacy_1994_H2_CO, MAPS_Zhang}, such that $\Sigma_{\mathrm{CO}} = \Sigma x$(CO). Therefore we can relate our derived vertical $^{12}$CO location ($z_{\mathrm{CO}}$) to the scale height of the disk and critical column density following,

\begin{equation}
    N_{\mathrm{CO}, \mathrm{crit}} = \frac{\Sigma_{\mathrm{CO}}}{\sqrt{2\pi}H}\int_{z_{\mathrm{CO}}}^{+\infty} \mathrm{exp} \left ( -\frac{z^2}{2H^2} \right) \,dz
\end{equation}

To solve the equation (4) we use a change in variable such that $z' = z/H$ and obtain an equation that can be solved numerically at each radial location where we have information on the $^{12}$CO emitting layer ($z_{\mathrm{CO}}$), to obtain the disk scale height if the emission is optically thick. 

\begin{equation}
    \int_{z_{\mathrm{CO}}/H}^{+\infty} e ^{z'^2/2} \,dz'= \frac{N_{\mathrm{CO}, \mathrm{crit}}\sqrt{2\pi}}{\Sigma_{\mathrm{CO}}}
\end{equation}

\begin{figure}[h!]
   \centering
   \includegraphics[scale=0.5]{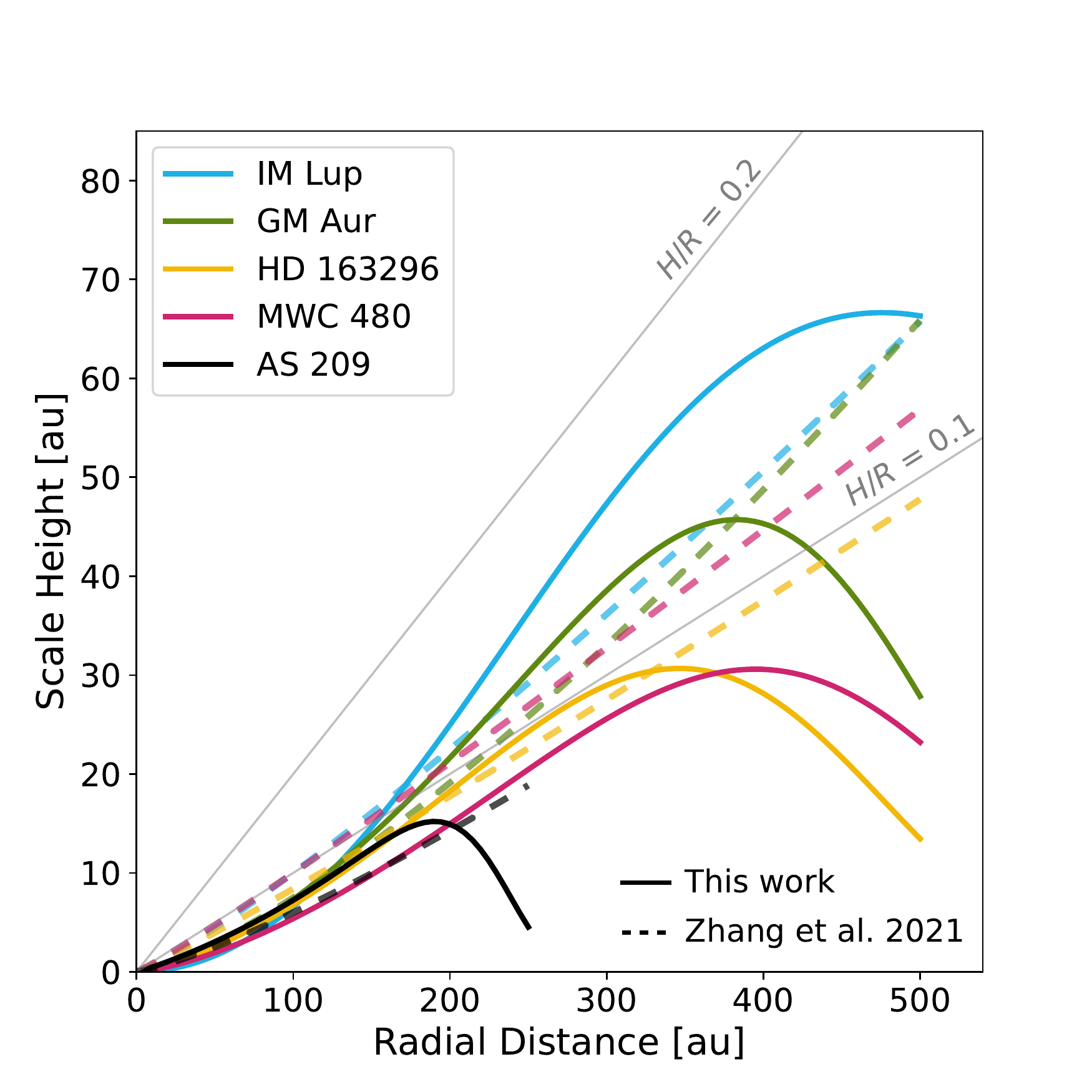}
      \caption{Derived gas pressure scale height ($H$) for each of the disks in the MAPS sample in solid colored lines from our analysis of the $^{12}$CO emitting surface. Dashed lines show the predicted scale height for each disk from \citet{MAPS_Zhang}. Grey solid lines show the location of constant $H/R$ at 0.1 and 0.2. 
              }
         \label{scale_height}
\end{figure}

Considering the surface density best-fit parameters derived from models in \citet{MAPS_Zhang}, we obtain the scale height profiles from our measurements for the disks in the MAPS sample. We note that the CO depletion profiles from \citet{MAPS_Zhang} are not considered because they are obtained for C$^{18}$O and C$^{17}$O, therefore we would have to make assumptions on the isotope ratios to convert them into $^{12}$CO depletion profiles, adding an additional uncertainty \citep{Miotello_2014}. This analysis is not done for Elias 2-27 and WaOph\,6 as we do not have a model of the surface density and replicating the study of \citet{MAPS_Zhang}, which involves thermochemical modelling and radiative transfer, is beyond the scope of this work. Figure \ref{scale_height} shows our estimation for the scale height based on the best-fit parametrization of $^{12}$CO surfaces, compared to the scale height profiles from the best-fit models studied in \citet{MAPS_Zhang}. Our results are in agreement with the models and with the canonical assumption of $H/R$ = 0.1 for a standard irradiated disk beyond $\sim$100\,au \citep[see equation 5 in][]{Lodato_2019_H_R}. 

\begin{figure*}[h!]
   \centering
   \includegraphics[width=\hsize]{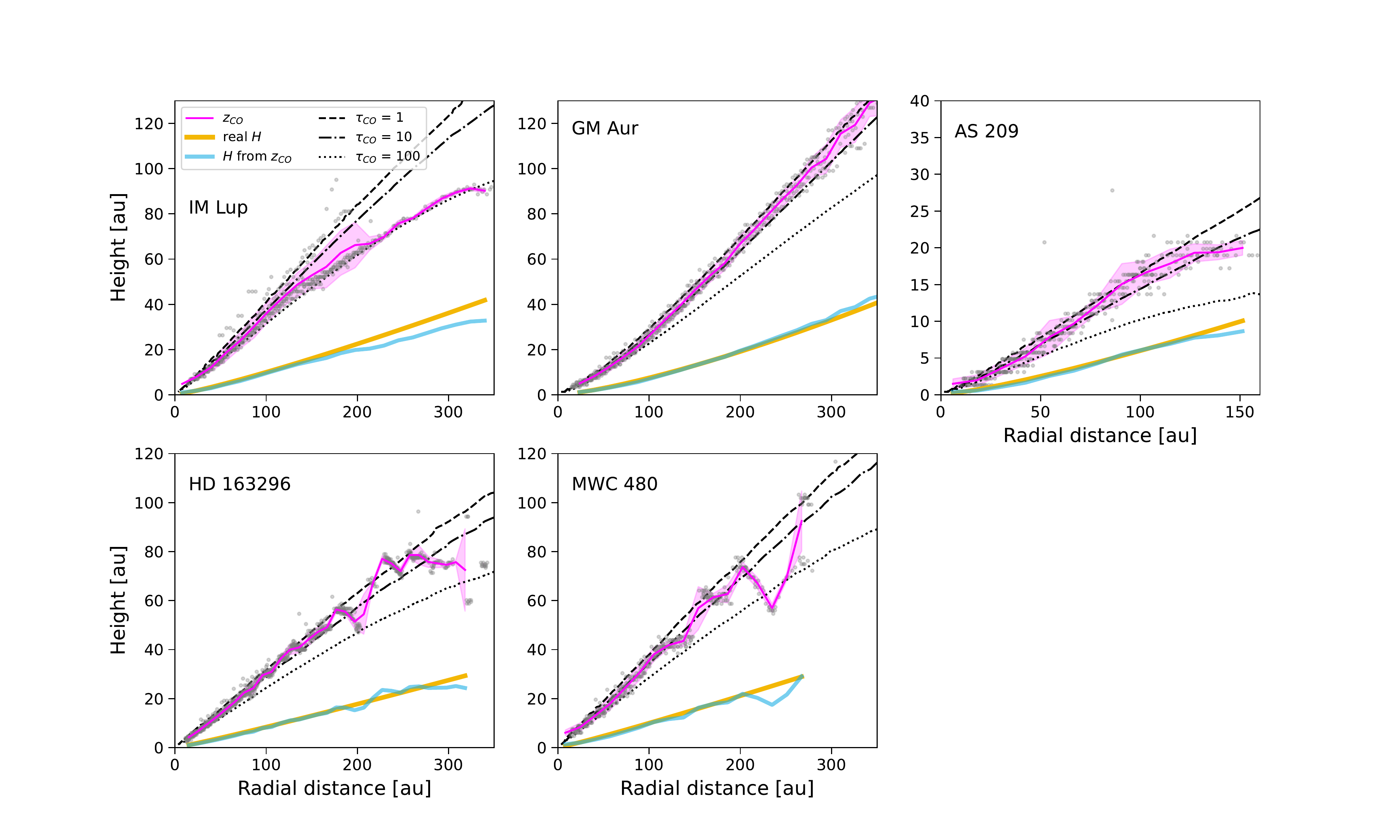}
      \caption{Results on the vertical location for $^{12}$CO (pink line) and the inferred pressure scale height (blue line). The $^{12}$CO vertical location is obtained with ALFAHOR through the analysis of mock channel maps generated from a DALI model which uses the surface density prescription of \citet{MAPS_Zhang} for each disk. Points extracted from the synthetic channel maps are shown in grey for comparison with the average value ($z_{\mathrm{CO}}$). The inferred pressure scale height is calculated following equation 5. The pressure scale height used as input for the DALI models ($H$) is shown for comparison (yellow line). Dashed, dot-dashed and dotted lines show the model location of the CO millimeter optical depth ($\tau_{CO}$) 1, 10 and 100, respectively. 
              }
         \label{DALI_test}
\end{figure*}

For IM Lup and HD\,163296 there are also estimates on the pressure scale height from the location of the scattered light surfaces \citep{ginski_2016_97048, Avenhaus_2018, Rich_2021}. Our values for the gas pressure scale height are below these estimates in both cases.  The results show a layering where the $^{12}$CO emitting surface is the most vertically extended, followed by the scattering surface and then the gas pressure scale height (see Fig. \ref{comp_hd16_imlup_height}). The latter is in agreement with expectations from theoretical models, where the gas pressure scale height is estimated to be 3-4 times lower than the scattering surface \citep{chiang_2001_height, ginski_2016_97048}. Our results display a difference of only 1.5-2 between the scattering surface and the gas pressure scale height. We note that the estimates we obtain of the pressure scale height strongly depend on the assumed density profile and CO abundance ($x(\mathrm{CO})$). Low densities ($\Sigma_{\mathrm{CO}}$ below or close to the assumed CO critical density) push the pressure scale height to large values, up to a factor of a few, with respect to the derived CO emission layer. This indicates that if we would consider a depletion factor in the CO abundance then the $H/R$ value would increase.

To test the robustness of our method we perform 2D thermochemical models with DALI \citep{Bruderer_DALI_2013}, using the gas surface density prescription of \citet[][]{MAPS_Zhang} and creating mock channel maps to retrieve the $^{12}$CO $J = 2-1$ vertical profile ($z\mathrm{_{CO}}$) for applying our methodology. The model details can be found in Appendix C.3 and the results of our test are displayed in Figure \ref{DALI_test}. We observe that, as expected in our model, the emitting layer we retrieve for $^{12}$CO from the mock channel maps is located at a similar height than the CO optical depth of $\tau_{CO}\geq1$. The optical depth value is an output of the DALI models and is calculated at each radial point by vertically integrating towards the disk midplane the sum of line and dust opacities and then subtracting the dust only integrated opacity value. In some cases the measured vertical profile for $^{12}$CO emission traces a location closer to the disk midplane, therefore it traces an optically thicker region reaching values of $\tau_{CO}\sim100$. This difference in the expected location of the $^{12}$CO emission does not seem to be related to any projection effects, such as disk inclination (see Figure \ref{DALI_inc}). IM\,Lup is the system that traces higher $\tau_{CO}$ in the outer disk and is also the source with the highest disk mass and radial extent (see Table \ref{table_DALI_param}). The resulting synthetic channel maps in this system are harder to trace in the outer radii than for the other sources, due to sudden brightness variations in the emission. The effect of this difficulty is reflected by the spread in the data points at 150-200\,au and low sampling further out (see Figure \ref{DALI_test}). Additionally, HD\,163296 and MWC\,480 show features in the model vertical structure that seem like the modulations studied in the previous section, however, it is uncertain if this is an actual physical effect or a structure that appears due to the coarser model resolution in the outer disk region. A future study focused on thermochemical models will aim at understanding the parameters that affect the exact location of the emission. The models shown in this section are just first approximations to future theoretical work and are not analyzed in detail here.

Using the extracted vertical profiles from the models, we apply our method to obtain the pressure scale height and recover an almost exact match to that of the scale height used as input for each source (see Figure \ref{DALI_test}). Differences between the inferred pressure scale height and the model input value are most apparent in the outer regions (r$\geq$250\,au), where the emission comes from more optically thick zone (most apparent in IM Lup as mentioned before, see Figure \ref{DALI_test}). A plot relating the conversion factor between the $^{12}$CO emitting layer and the gas pressure scale height, for different $\tau_{CO}$ values can be found in Appendix C. 

\begin{figure*}[h!]
   \centering
   \includegraphics[scale=0.7]{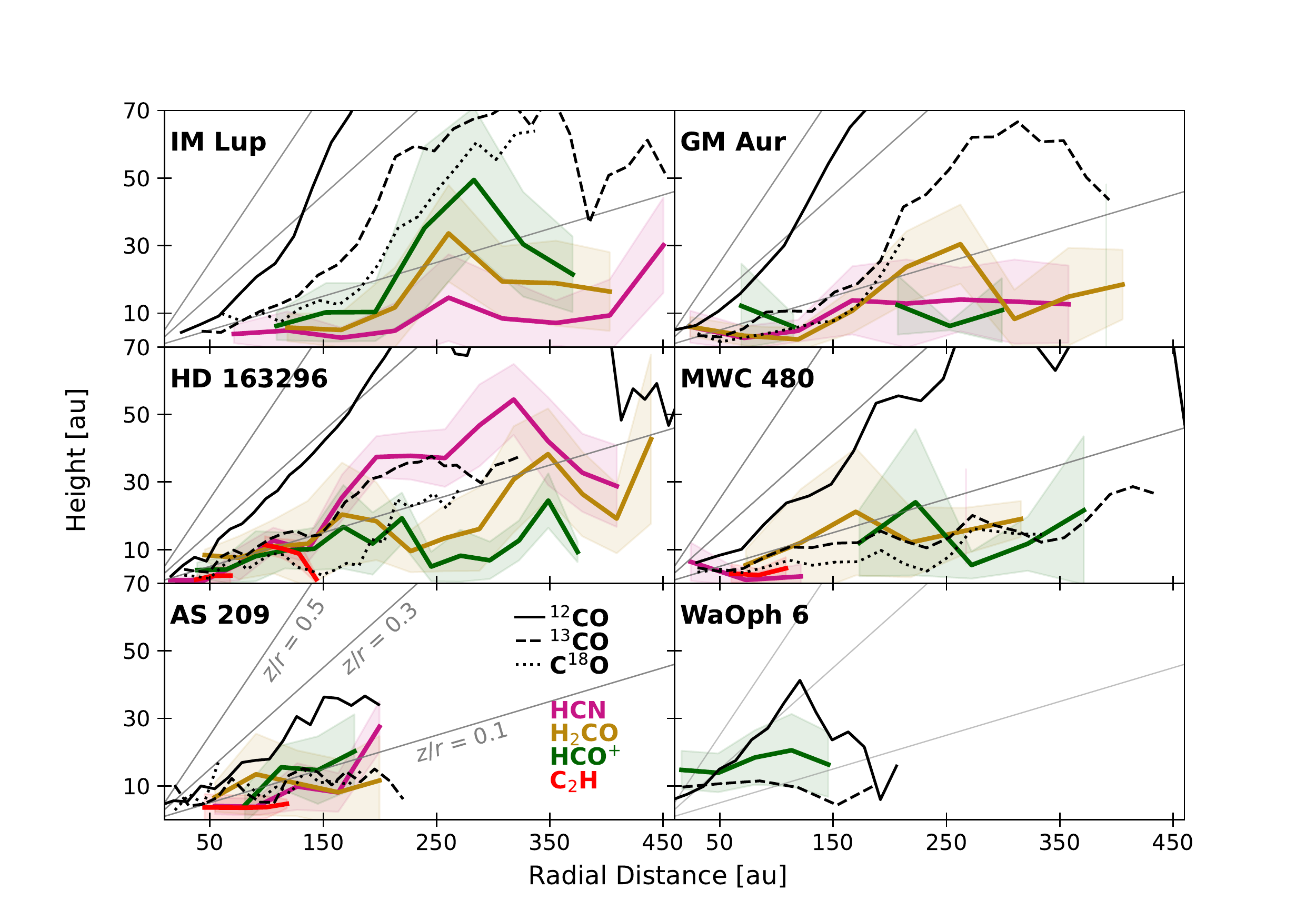}
      \caption{Vertical profiles of the emission surface for molecules other than CO isotopologues in the studied disks. Elias 2-27 is not shown due to lack of data in other tracers. Black curves show the location of the CO isotopologues in $J = 2-1$ transition for reference. Solid line is $^{12}$CO, dashed line $^{13}$CO and dotted line is C$^{18}$O. Each colored curve shows a different tracer, where the solid line is the mean value and the shaded region shows the dispersion of the retrieved data points within each radial bin. Grey lines mark constant $z/r$ of 0.1, 0.3 and 0.5.
              }
         \label{panel_other_molec}
\end{figure*}

While this observationally-driven derivation of $H$ should be considered as an approximate first trial on describing the disk pressure scale height using the location of CO emission, our initial test with thermochemical models show it is a good approximation. We expect this method to hold if the gas in the disk follows Keplerian rotation, $^{12}$CO emission is optically thick and the vertical distribution is similar to a Gaussian profile. Under these assumptions, our estimate shows that all systems have scale heights between $H/R$ of 0.1-0.2 (Figure \ref{scale_height}), which is in agreement with theoretical predictions. An important caveat to our comparison with DALI models is that we have tested our method (equation 5) with a model that uses a Gaussian distribution for the vertical profile (see Appendix C for details), which is our initial assumption for inferring the pressure scale height (equation 3). While our tests are an important check to assure that our method is consistent, future work must focus on the errors that may appear if the vertical density profiles deviate from a Gaussian profile and consider an analytical disk prescription that gives a better representation of the vertical temperature structure \citep[e.g. ][]{Chiang_Goldreich_1997, Dullemond_2002}.

\subsection{Other tracers}

For the MAPS program targets we are able to trace the vertical profile of HCN, H$_2$CO, HCO$^+$ and C$_2$H in the same way as with the CO isotopologues. In the case of WaOph\,6, the lack of spectral and spatial resolution in the data only allowed us to recover the surface of HCO$^+$, even though HCN and H$_2$CO emission data is available too (see Appendix A). The vertical profiles of these molecules are shown, compared to the CO vertical layers, in Figure \ref{panel_other_molec}. Elias 2-27 has observations of CN $N=3-2$ hyperfine transitions from which the emitting surface has been recovered \citep{Paneque-Carreno_2022_Elias_CN}, however, as we are able to recover the emitting surface of CN $N=1-0$ only in HD\,163296 (shown in Figure \ref{panel_hd16}), we do not display this molecule in Figure \ref{panel_other_molec}. 

\begin{figure*}[h!]
   \centering 
   \includegraphics[scale=0.42]{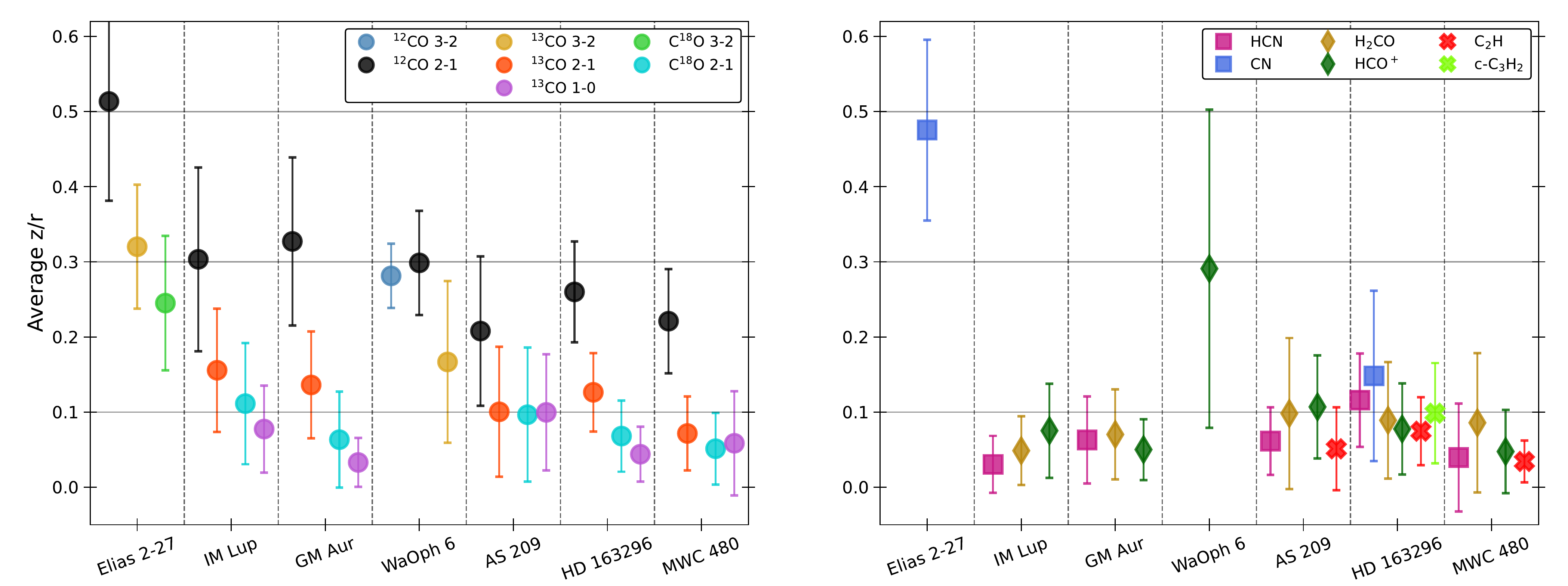}
      \caption{Average $z/r$ values of each tracer and disk under study. To avoid the effect of the turnover in the case of vertically extended molecules, only data within 80\% of the $^{13}$CO (or $^{12}$CO for MWC\,480 and AS\,209) r$_{taper}$ is considered. Note that for Elias 2-27 CN is a higher transition ($N = 3-2$) than the CN shown in HD\,163296 ($N = 1-0$).
              }
         \label{z_over_r}
\end{figure*}

\subsubsection{Complete sample}

Molecules beyond CO isotopologues seem to mostly reside close to the midplane. Due to this, we are not able to visually separate upper and lower sides of the disk when masking the channel maps to extract the vertical profiles. This implies that there could be some contamination and scatter caused by the lower side of the disk. Additionally, for some tracers the vertical extent of the emission is comparable with the beam size, therefore the profiles of these molecules are considered tentative results. Higher spatial resolution and better SNR data may allow us to constrain them in more detail in the future. The gaps in the vertical profiles are due to lack of data (under two points) in the radial bin (see Method section). We note that our results are in agreement with those obtained by \citet{MAPS_Law_Surf} for HCN and C$_2$H in the inner $\sim$150\,au. 

\begin{figure*}[h!]
   \centering
   \includegraphics[width=\hsize]{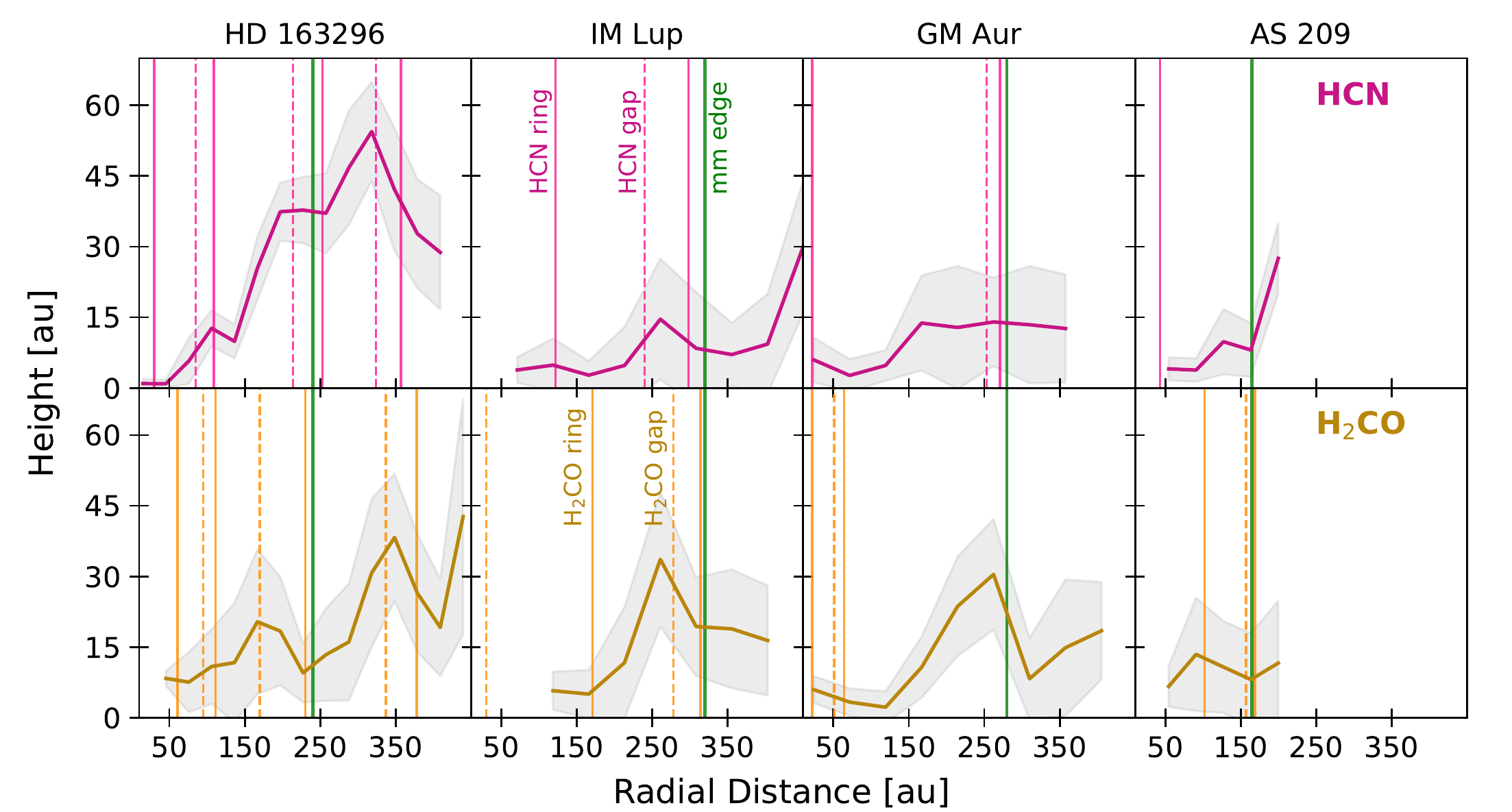}
      \caption{Vertical surface profiles for HCN in the top row and H$_2$CO in the bottom row. Green vertical line marks the edge of the dust continuum as reported in Law et al. 2020a. Solid and dashed lines in each panel mark rings and gaps, respectively found in the emission maps of each molecule from Law et al. 2020a. In the profiles, solid line marks the mean value and grey shaded area the dispersion of the retrieved data points in each radial bin.
              }
         \label{panel_H2CO_HCN}
\end{figure*}

The overall average $z/r$ values of each tracer are shown in Figure \ref{z_over_r}. As various tracers have different turnover radii and extent, we follow a similar approach to \citet{Law_2022_12CO} and consider the emission only from within 80\% of the $^{13}$CO $J = 3-2$ r$_{taper}$. For MWC\,480 and AS\,209 we use $^{12}$CO, as no turnover is detected in $^{13}$CO (see Table \ref{table_vertical_co}). The emission from most tracers beyond CO isotopologues has a large scatter, but overall comes from regions with $z/r \leq$0.1 (see right panel of Fig. \ref{z_over_r}). This is in agreement with the flat morphology of the channel maps and the assumptions of flat disk used in MAPS for emission other than CO \citep[e.g.][]{MAPS_Law_radial, MAPS_Guzman, MAPS_Bergner}. HCN rises from an upper layer only in HD 163296 and HCO$^{+}$ shows a moderate elevation in IM Lup, AS 209 and WaOph 6. H$_{2}$CO has a consistent behaviour across all stellar masses, with $z/r \sim$0.1. The emitting surface of CN can only be traced in Elias 2-27 and HD\,163296. Both disks show CN is the highest emitting molecule, aside from $^{12}$CO, however in Elias 2-27 it traces the same region and in HD\,163296 it is in a middle layer closer to the rest of the molecular reservoir.

The stars in Figure \ref{z_over_r} are ordered in the horizontal axis by increasing stellar mass, however the scale is not linear as several of them have almost equal mass value (see Table \ref{table_sample_all}). A tentative relation between the $z/r$ values of CO tracers and stellar mass may be recovered from the left panel of Figure \ref{z_over_r} such that $z/r$ decreases with increasing stellar mass. This relationship has been tested for $^{12}$CO \citep{MAPS_Law_Surf, Law_2022_12CO} and in this work we tentatively recover it for $^{13}$CO $J=2-1$ and C$^{18}$O $J = 2-1$ too (orange and light blue dots in Fig. \ref{z_over_r}). Contrary to the behaviour of its higher transitions, $^{13}$CO $J=1-0$ emits from a lower $z/r \leq$0.1 layer and does not have any correlation to the stellar mass. Considering the mean values we note that $^{13}$CO $J=2-1$ lies in a vertical region 2-3 times lower than $^{12}$CO $J=2-1$ for all the disks in the MAPS sample. This also applies for WaOph\,6 in $J=3-2$ transitions, but for Elias 2-27 $^{13}$CO is located closer to $^{12}$CO and the ratio between their mean values is 1.5.

\subsubsection{Structure in HCN and H$_2$CO}

The HCN and H$_2$CO vertical profiles are well sampled (see all of the extracted data points in Appendix B) and show distinct morphologies, that in most cases do not follow an exponentially tapered power-law or a single power-law model. These profiles are presented separately in Figure \ref{panel_H2CO_HCN} for all disks in the MAPS sample except MWC 480, which is not included due to the lack of data points for HCN emission. Overlayed to the profiles are the locations of rings and gaps found in the integrated emission of each corresponding molecule and the edge of the millimeter continuum \citep{MAPS_Law_radial}. 

HCN shows a step-like profile in HD 163296, with peaks that mostly coincide with gaps in the integrated HCN emission. IM Lup shows a peak in the HCN profile at $\sim 270$\,au, which is in between a gas gap and ring and also coincides with the single peak seen in the H$_2$CO vertical profile. GM Aur and AS 209 may have tentative HCN peaks around $\sim 150$\,au, however the data do not allow us to properly characterize the profiles and there is a large dispersion in the recovered data points from the channel maps. While HCN is expected to trace high layers in the disk atmosphere \citep{Visser_2018, MAPS_Bergner}, it is clear that this is only the case in HD 163296 (see also Fig. \ref{z_over_r}).

The H$_2$CO vertical profiles can be traced with less scatter in all disks and for HD 163296, IM Lup and GM Aur we see clear peaks. In all cases there seems to be a peak close to and just inwards with respect to the edge of the millimeter continuum. The peaks in the vertical profile of HD 163296 and IM Lup coincide with gas gaps, as was the case for HCN. GM Aur has gas rings and gaps detected only in the molecular emission at small ($<$100\,au) radii, however it has a strong, well constrained rise in the emitting layer close to the continuum edge.

In HD 163296 and IM Lup there seems to be a correlation between the presence of line emission gaps and the peaks in the vertical structure for both HCN and H$_2$CO. It could be that the substructure of the molecular emission affects our profiles, as we are tracing the data directly from the channel maps, assuming that the emission maxima trace an isovelocity curve. However, we do not see this effect in GM Aur, where H$_2$CO also has a strong vertical perturbation and we note that most of the gaps in the molecular emission at outer radii in HCN and H$_2$CO have very low-contrast \citep{MAPS_Law_radial}, therefore we do not expect them to have such a noticeable impact on the retrieved profile. All vertical profiles show peaked features beyond the continuum emission border, therefore it is unlikely that this is an effect of continuum over-subtraction either. Overall, if the emission is optically thin, which is expected from the brightness temperature profiles of HD\,163296 (Fig. \ref{panel_hd16}) and previous studies \citep{MAPS_Bergner, MAPS_Guzman}, the vertical profiles seem to be tracing distinct variations of the physical location from where HCN and H$_2$CO are being emitted.

\section{Discussion}

\subsection{CO isotopologue vertical layering}

There have been dedicated studies to predict the regions from where molecules emit in protoplanetary disks \citep{van_Zadelhoff_2001_vert_models, Aikawa_2002, Walsh_2010, Miotello_2014, woitke_2016_prodimo}. We trace the location of the emitting region for various CO isotopologues in our seven disks, including different transitions of $^{13}$CO in the MAPS sample and $^{12}$CO in WaOph 6. Our analysis shows that $^{12}$CO emission comes from a wide range of vertical locations, which is in agreement with previous studies \citep{MAPS_Law_Surf, Law_2022_12CO}. This does not correlate directly with the emission location of other CO isotopologues or different molecules, except possibly $^{13}$CO (see Section 3.3 and Figure \ref{z_over_r}). 

We take as reference the work by \citet{van_Zadelhoff_2001_vert_models}, where literature disk models are processed using a full 2D Monte Carlo radiative transfer code to determine the vertical location of the emitting surfaces for CO and HCO$^{+}$isotopologues. The locations traced for the CO isotopologues in our work are qualitatively compared with the expected location of CO for a chemical abundance similar to the interstellar medium, and a more depleted abundance. In both scenarios it is expected to have a layering where $^{12}$CO traces the upper layers, followed by $^{13}$CO and C$^{18}$O, considering all transitions \citep[see also ][]{Miotello_2014}. For the disks in the MAPS sample, we clearly recover that $^{12}$CO emission comes from a higher region than $^{13}$CO. Also, for $^{13}$CO the $J=2-1$ transition emission surface is above the $J=1-0$ transition, as expected. However, we  observe that C$^{18}$O $J = 2-1$ is emitting from a layer very close to $^{13}$CO $J = 2-1$ and in most cases above $^{13}$CO $J = 1-0$, which is not expected from the models. This overlap in the emission regions of C$^{18}$O $J = 2-1$ and $^{13}$CO $J = 1-0$ could be explained by both of them emitting from just above the CO freeze-out region, which is also coherent with the measured low brightness temperature (see Fig. \ref{panel_temp}). Indeed, it has been proposed that C$^{18}$O $J = 2-1$ emission in IM Lup traces the freeze out \citep{Pinte_2018_method}. Our new results showing that $^{13}$CO $J = 1-0$ emits from the same region or even closer to the midplane may be additional evidence of the freeze-out region location.  C$^{18}$O $J = 1-0$ is also detected for the MAPS sample but unfortunately, due to the low SNR of this transition, it is not possible to trace its emitting surface, which would be useful for additional comparison.

In WaOph 6 $^{12}$CO $J = 3-2$ and $J = 2-1$ trace the same region, which is in agreement with the previously discussed theoretical models. Future work obtaining observations at higher transition such as $^{12}$CO $J = 6-5$ could help differentiating between a depleted or ISM-like abundance, depending on how similar the emitting region is to that traced by lower transitions \citep{van_Zadelhoff_2001_vert_models}. Elias 2-27 shows a highly elevated emission layer, where the $J = 3-2$ transitions of $^{13}$CO and C$^{18}$O are very close to the  $^{12}$CO $J = 2-1$. These two disks do not have a detailed description of their surface density or temperature structure, which would be useful for comparing them with the rest of the sample and obtaining the gas pressure scale height. Both disks have spiral structures in the dust continuum emission, which have been proposed to be linked to the effects of the disk self-gravity \citep{DSHARP_Huang_Spirals, Paneque-Carreno_Elias1}. Our results show that in the line emission they both have very different structures both radially and vertically, therefore it remains unclear how dust continuum features relate with the vertical material distribution.

\subsection{Vertical distribution of other molecules}

Our work presents direct constraints of the emitting location for several molecules, extended to radial distances beyond $\sim$150\,au. Overall, the results show that most of these molecules emit from regions very close to the midplane ($z/r$<0.1, see Fig. \ref{z_over_r}), which conflicts with theoretical predictions, in particular when analyzing the location of UV sensitive tracers such as HCN and C$_2$H \citep{Aikawa_2002, Walsh_2010, MAPS_Bergner, MAPS_Guzman}.

Various physical-chemical models have predicted that CN and HCN emission should be sensitive to UV flux \citep{Cazzoletti_2018_CN, MAPS_Bergner}. The emitting layer of HCN should be located just below CN, due to the higher photodissociation rate of HCN compared to CN \citep{MAPS_Bergner}. Overall, the emission of both tracers is expected to arise from the upper atmosphere of disks, in regions close to the  $^{12}$CO emitting layer \citep[see also][]{Walsh_2010}. The only system where we are able to extract the emitting surfaces of both tracers is HD\,163296. In this disk we find that CN $J = 1-0$ emission is located in an intermediate layer, between $^{12}$CO and $^{13}$CO $J = 2-1$ (see Figs. \ref{panel_hd16} and \ref{z_over_r}). In all the other disks, except for Elias 2-27, it is not possible to extract CN surfaces and HCN traces a region close to the midplane ($z/r$<0.1). Observing emission from regions closer to the midplane may be an indicator of deeply penetrating UV flux either from the central star or from an external source \citep[see models by][]{Flores_2021}. Another alternative is that X-ray radiation is responsible for the emission, as suggested for the Flying Saucer system \citep{RuizRodriguez_2021}. UV radiation is deeply affected by the location of small dust particles, therefore the constraints on the emitting surfaces shown in this work may be used for dust growth and settling models in disks through simultaneous modeling of the dust location, UV radiation and molecular excitation. C$_2$H is also a molecule expected to be a good UV tracer \citep{Miotello_2019_c2h, MAPS_Guzman} and even though it is less radially extended, in the inner disk it traces the same region as HCN. Future observations of higher CN transitions may allow us to study the distribution of this molecule for the rest of the sample and compare its location with the retrieved HCN emitting region.

Models suggest that HCO$^+$ should be emitting from a region similar to the CO emission layer \citep{Aikawa_2002}. We find that the average $z/r$ of the molecule is indeed close to C$^{18}$O. In IM Lup, AS 209 and  WaOph 6 the emission comes from above $z/r$ of 0.1. These disks are expected to be the youngest of the sample \citep{DSHARP_Huang_Spirals, MAPS_Oberg}, which could imply they have had less chemical reprocessing and CO abundances similar to ISM. Disks in which CO has been transferred into other species (expected for older systems) have  HCO$^+$ emission located closer to the midplane, compared to models with ISM-like abundances \citep{van_Zadelhoff_2001_vert_models}. Constraints from the measured excitation temperature of H$_2$CO \citep{Pegues_2020} and chemical models \citep{Walsh_2010} also predict that it should be located in the molecular layer, but above the CO freeze-out region. Indeed, we trace the emission from H$_2$CO in a very similar region to that of C$^{18}$O and HCO$^+$.

The morphology of the HCN and H$_2$CO emitting layers show a distinct peaked structure (see Fig. \ref{panel_H2CO_HCN}). A possible relation is observed between the location of radial gaps in the molecular distribution and the peaks in the vertical profile. H$_2$CO is a molecule that has two formation pathways, it may emit due to desorption processes, from dust grains close to the midplane, or warm gas phase chemistry \citep{van_Scheltinga_2021_h2co, Carney_2017, MAPS_Guzman}. The distinct surfaces we trace in H$_2$CO could be the first direct indication of regions where different emission mechanisms are acting. In particular, modeling of HD\,163296 by \citet{Carney_2017} predicts that in order to reproduce the radial intensity profile the outer disk (beyond the millimeter edge) must have an increased H$_2$CO abundance. The model considered a higher abundance compared to that of the inner disk that extends vertically \citep[see Fig. 5 of][for the R-step case]{Carney_2017}, which is coincident with the peak traced in the emission surface of H$_2$CO in HD\,163296 beyond the millimeter edge (see Fig. \ref{panel_H2CO_HCN}). It is important to note that the results of \citet{Carney_2017} showed that the H$_2$CO could not be produced uniquely by gas-phase reactions. Our vertical profile shows that at radius $\sim$240\,au the emission comes from closer to the midplane, which would be coherent with thermal desorption of the molecule from icy grains.

Previous works focused on H$_2$CO emission have not recovered the structure we see in this study, however, models for DM Tau considering gas-phase and grain chemistry recover a bulged H$_2$CO density structure in the vertical profile \citep{Loomis_2015}. This bulged feature is expected at a higher elevation than what we trace in our vertical profiles and may not necessarily be reproduced in the emission surface. In a study by \citet{Pegues_2020} it was found that measurements of the H$_2$CO excitation temperature in various systems show mostly constant radial temperatures, however clear peaks are detected in LkCa15 and J1604-2130. If the densities are sufficient to assume that the emission is in Local Thermodynamic Equilibrium (LTE) then the excitation temperature may be a proxy to the kinetic temperature and thus to the vertical emission distribution. For HCN there are no sources in the literature that show evidence of a structured vertical distribution and the excitation temperature profiles derived for the MAPS sample do not have any noticeable relation to the features seen in our vertical profiles \citep{MAPS_Guzman}.

\subsection{CO modulations; correlation to millimeter gaps and kinematic features}

Vertical structure is found in the CO isotopologue emission for IM Lup, HD163296 and MWC 480, this structure can be modelled as drops from a determined baseline, using gaussian functions (see Section 3.2.1) . We name these drops modulations. There seems to be a consistent relation between the locations of the modulations, gaps observed in millimeter continuum and kinematical residuals. As the emission surfaces are traced directly from the channel maps there are several possibilities regarding their origin. It could be that, if the kinematical features are caused by a planetary companion, the vertical modulations are a tracer of the material infalling and possibly accreting on to the planet. The kinematic residuals used in this work have all been related with the presence of planetary companions \citep{Pinte_2018_hd16planet, Teague_2018_hd16planet, Izquierdo_2021_hd16planet, MAPS_Teague}. If there are planets in these disks, it has been shown that meridonial flows of material \citep{Fung&Chiang_2016} will be capable of disturbing the vertical density structure \citep{Morbidelli_2014, Szulagyi_2017, Szulagyi_2022}. Meridonial flows have been detected observationally in HD\,163296 \citep{Teague_2018_hd16planet}, however, it is unclear if the modulations we detect in the vertical disk structure are related to the meridonial flows, as simulations do not provide clear predictions on the vertical structure, for they are mainly focused on density and velocity perturbations. It is expected that the vertical profiles trace the location where the gas becomes optically thick, therefore the modulations could also be related to density drops. However, as mentioned in section 3.2.1, the locations of vertical modulations do not coincide with the gaps and rings in the integrated emission map of each molecule \citep{MAPS_Law_radial}, which should be a direct proxy for density variations.

Another possibility could be that there are no real vertical modulations, but rather we are recovering artifacts as we assume that the emission maxima trace an isovelocity curve in a smooth disk, but in the sample there are perturbed channel maps \citep[MWC 480, HD 163296, ][]{MAPS_Teague}. Alternatively, it could also be that the emitting layer is indeed perturbed and causing the channel map to appear perturbed due to projection effects. Future detailed studies on the effect of velocity perturbations in the extraction of emission surfaces will aid in breaking this degeneracy and understanding the observational characteristics of each effect. However, as we clearly recover the modulations at a constrained radial distance on both sides of the disk, with respect to the semi-minor axis and in all of the studied channels, we propose that they are real perturbations present in the vertical structure.

If the surface modulations are related to planetary companions in the disks, the feature at $\sim$393\,au in IM Lup is particularly relevant. It is a coherent, strong and well defined modulation present in $^{12}$CO and $^{13}$CO emission, with no kinematical feature associated to it. The radial location of this dip is close to where the gas component of the disk is expected to have a sharp change in the surface density \citep{Panic_2009}. Searches for planetary companions in IM Lup have been performed \citep{Mawet_2012, Launhardt_2020} but no object has been detected, considering a detection probability of $\sim$30\% for a planet of 13M$_J$ at 400\,au. For MWC 480 and HD163296 there are coherent inner modulations that are coincident with dust gaps, which for MWC 480 have no kinematical counterpart, however for HD163296 there is a strong linewidth residual at $\sim$38\,au and a positive velocity gradient at $\sim$45\,au \citep{Teague_2018_hd16planet, Izquierdo_2021_hd16planet}. Detailed models studying the effect of planetary companions on the CO emitting layer will be useful to determine if the detection of vertical features may be an indirect measure of the presence of a planet, but this goes beyond the scope of this paper.

\section{Conclusions}

In this work we have presented observational constraints on the vertical location of the molecular emission of seven disks. Using data of high angular and spectral resolution we directly trace the emission location for up to ten different molecules. Our main findings and conclusions are the following:

\begin{enumerate}
    \item We have characterized the emission surfaces of multiple molecules through a geometrical analysis of the channel map emission. We have also detected structured emission layers with modulations and spikes. Using an implementation of the \citet{Pinte_2018_method} method, where we mask the observed emission layer, we trace the emission surfaces at larger radii and for lower SNR molecular emission compared to previous work.
    
    \item The derived emitting surfaces for CO isotopologues show a clear layering, with $^{12}$CO being the most vertically extended tracer. $^{13}$CO and C$^{18}$O  of the same transition trace similar regions. From our available data we can also corroborate that lower transitions trace closer to the midplane than higher transitions in the case of $^{13}$CO, consistent with thermochemical models. 
    
    \item If the emission is optically thick and good approximations for the surface density and CO abundance are available, the $^{12}$CO emission surface can be used to obtain the gas pressure scale height of the disk. Our values indicate that the scale height is consistent with values of $H/R = $0.1. These results are obtained using a simple analytical one layer disk model and initial testing with thermochemical DALI models indicate the method is robust, at least if the vertical material in the disk follows a Gaussian distribution.

    \item The locations of the detected modulations in the CO isotopologue emitting surfaces are correlated to that of millimeter continuum gaps and kinematic perturbations. These modulations may be related to the presence of planetary companions or other mechanisms causing variations in the gas density distribution.
    
    \item HD\,163296 shows a rich molecular reservoir from which most tracers can be located in distinct vertical regions. Overall in our sample most of the molecular emission for tracers other than CO originates close to the midplane at $z/r$<0.15. In the case of HCO$^+$ and H$_2$CO the observational constraints agree with theoretical predictions and constraints on the excitation temperature. Both molecules seem to be tracing the molecular layer above the CO freeze-out region. The particular morphology of the H$_2$CO emitting surface may be the first direct indicator of this molecule originating from both ice grain desorption and gas phase chemistry.
    
    \item HD 163296 displays CN and HCN in an intermediate vertical region. Their location is in agreement with theoretical predictions \citep[e.g.][]{Cazzoletti_2018_CN, MAPS_Bergner} about HCN tracing a layer just below CN. The rest of the systems in the MAPS sample show HCN very close to the midplane and it is not possible to retrieve CN emission surfaces to compare with. This is not expected from theoretical models. Future observations of higher CN transitions will allow us to compare our results on the location of HCN and better understand the disk radiation conditions. 
    
    \item This sample of disks and molecules represent the largest survey to date on the direct characterization of the emitting regions for multiple tracers. Dedicated chemical-physical modelling is crucial to understand the diversity in the location of each molecule and how to relate the vertical profiles to actual disk properties.
\end{enumerate}

We aim for this work to be used as a reference catalogue for future dedicated models on each of the sources so that the chemical-physical origin of each emission line can be adequately studied. With the sensitivity of instruments like ALMA we hope to enlarge the sample of disks where this kind of study is possible.

\begin{acknowledgements}
We thank the referee for the constructive comments. This paper makes use of the following ALMA data: \#2015.1.00168.S, \#2016.1.00484.L, \#2016.1.00606.S,  \#2017.1.00069.S and \#2018.1.01055.L. 
ALMA is a partnership of ESO (representing its member states), NSF (USA), and NINS (Japan), together with NRC (Canada),  NSC and ASIAA (Taiwan), and KASI (Republic of Korea), in cooperation with the Republic of Chile. The Joint ALMA Observatory is operated by ESO, AUI/NRAO, and NAOJ. 
Astrochemistry in Leiden is supported by the Netherlands Research School for Astronomy (NOVA), and by funding from the European Research Council (ERC) under the European Union’s Horizon 2020 research and innovation programme (grant agreement No. 101019751 MOLDISK).

\end{acknowledgements} 

\bibliographystyle{aa}
\bibliography{vert_paper.bib}

\begin{appendix}

\section{WaOph 6 data and self calibration}

The emission lines studied in WaOph\,6 are mostly observed by ALMA program \#2015.1.00168 (P.I. G. Blake). The data was obtained in two consecutive observations, performed on the 28 of June 2016 and the total integration time on source was 29 minutes. Table \ref{waoph6_data} shows the information for all of the detected lines together with their integrated emission and imaging parameters after self-calibration. Besides the detected species, the spectral setup was also intended to observe H$^{13}$CN $J = 4-3$ and SO$_{3}$ $7(8) - 6(9)$. We do not detect either of them even when using a natural weighting of the visibilities, at the achieved sensitivity level of 6mJy\,beam$^{-1}$ (beam values for these tracers are $\sim$0.4$\arcsec$). Of the detected lines, not all of them can be used to extract the emitting surfaces, due to low SNR, compact radial extent and lack of spatial resolution. Of this dataset only $^{12}$CO, $^{13}$CO and HCO$^{+}$ have their emission surfaces extracted in the main text.

The data was initially calibrated the ALMA pipeline. Afterwards phase and amplitude self-calibration were performed to enhance the signal to noise of the emission. To do this we use the tools found in the Common Astronomy Software Applications \citep[CASA, ][]{McMullin_CASA}, version 6.1.1.15. The self calibration solutions were found using the line-subtracted continuum channels in all available spectral windows. Solutions for phase and amplitude were considered until the SNR improvement of the continuum was less than 2\%. Four rounds of phase self-calibration were applied, using a maximum time interval of 965 seconds and dividing by two in each iteration to reach a minimum time interval of 120 seconds. After this, two rounds of amplitude self-calibrations were done, using a solution time interval of 965 and 480 seconds in each iteration. The overall peak signal to noise improvement was of 1300\% in the continuum after completing self-calibration.

\begin{figure*}[h!]
   \centering
   \includegraphics[width=\hsize]{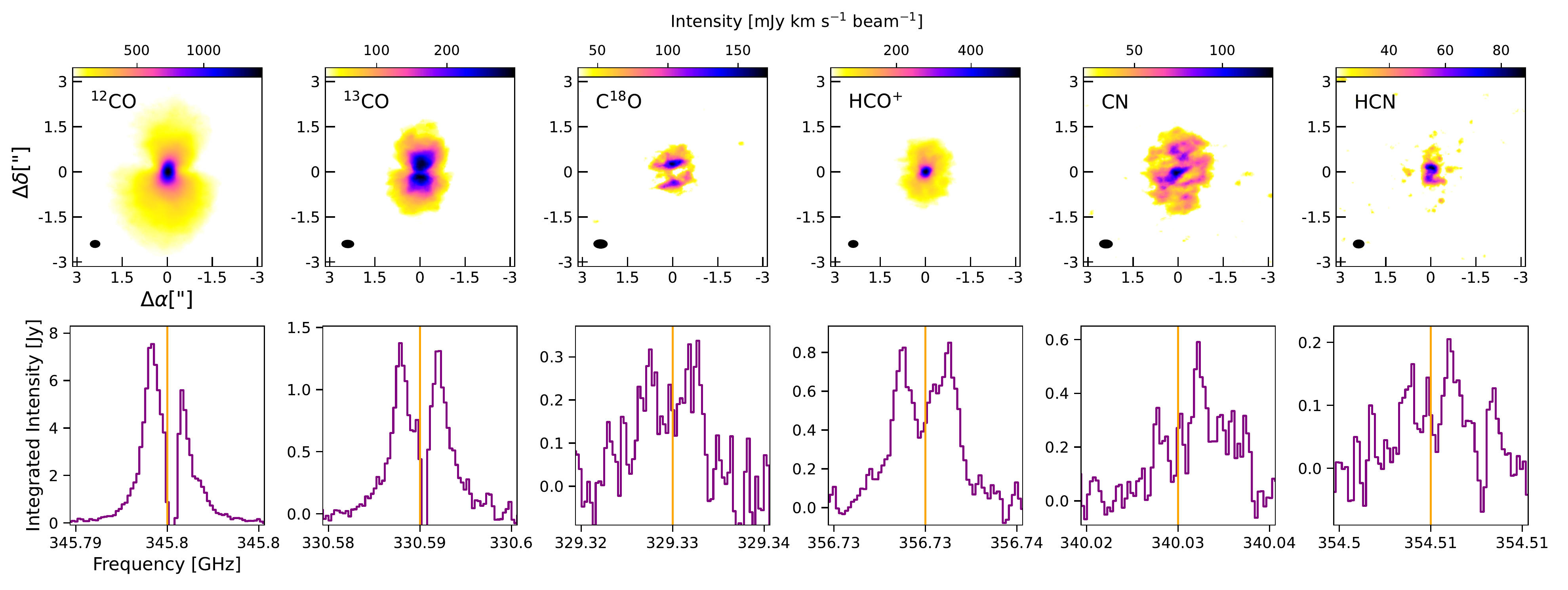}
      \caption{For each of the detected molecules in WaOph\,6 from program \#2015.1.00168.S (P.I. G. Blake), top row displays the integrated emission maps (moment 0) with an emission cutoff at 3$\sigma$. Bottom row shows the emission spectrum of each line, vertical orange line indicates the natural frequency of the molecule.
              }
         \label{waoph_emission}
\end{figure*}

In the self-calibrated continuum data the center of emission was found through a Gaussian fit in the image plane. Both phase and pointing centers in the whole dataset were adjusted using CASA tasks \textsc{fixvis} and \textsc{fixplanets} respectively. The self-calibration solutions are then applied to the line emission channels and images of each line are produced using multiscale \textsc{tclean} and varying robust parameters to compromise between SNR and spatial resolution. Details on the robust and final FWHM beam values are found in Table \ref{waoph6_data}. In all cases a stopping threshold of 3$\sigma$ is used to compute the image. The integrated emission (moment 0) maps and line spectra are shown in Figure \ref{waoph_emission}.

\begin{table*}[h]
    \centering
    \def\arraystretch{1.3}
    \setlength{\tabcolsep}{4pt}
    \caption{Imaging and emission parameters for WaOph\,6 data.}
    \begin{tabular}{c| c| c |c | c | c c c}
    \hline 
         Molecule & Transition & Freq. & Int. Flux & Robust & &Beam& \\
          &  & [GHz] & [Jy\,km\,s$^{-1}$] &  & $b_{min}$&  $b_{max}$ & $PA$ \\
    \hline 
         $^{12}$CO & $J = 3-2$ & 345.795 & 21.343 $\pm$ 0.069 & 0.5 & 0.27\arcsec & 0.35\arcsec & 86.77$^{\circ}$\\
         $^{13}$CO & $J = 3-2$ & 330.588 & 5.073 $\pm$ 0.069 & 0.5 & 0.28\arcsec & 0.43\arcsec & -85.74$^{\circ}$ \\
         C$^{18}$O & $J = 3-2$ & 329.330 & 1.364 $\pm$ 0.087 & 0.5 & 0.31\arcsec & 0.48\arcsec & -80.11$^{\circ}$ \\
         HCO$^+$ & $J = 4-3$ & 356.734 & 3.988 $\pm$ 0.077 & 2.0 & 0.26\arcsec & 0.35\arcsec & 87.05$^{\circ}$ \\
         CN & $J = 5/2-3/2$ & 340.031 & 2.329 $\pm$ 0.064 & 2.0 & 0.3\arcsec & 0.46\arcsec & -80.59$^{\circ}$ \\
         HCN & $J = 4-3$ & 354.505 & 0.747 $\pm$ 0.071 & 2.0 & 0.3\arcsec & 0.39\arcsec & 85.77$^{\circ}$ \\
    \hline
    \end{tabular}
    \tablefoot{This table shows only the data from program \#2015.1.00168.S (P.I. G. Blake).
    }
    \label{waoph6_data}
\end{table*}

\section{Data and best-fit parameters for the emitting layers}

\subsection{Data points and averaged vertical profiles}

\begin{figure*}[h!]
   \centering
   \includegraphics[scale=0.65]{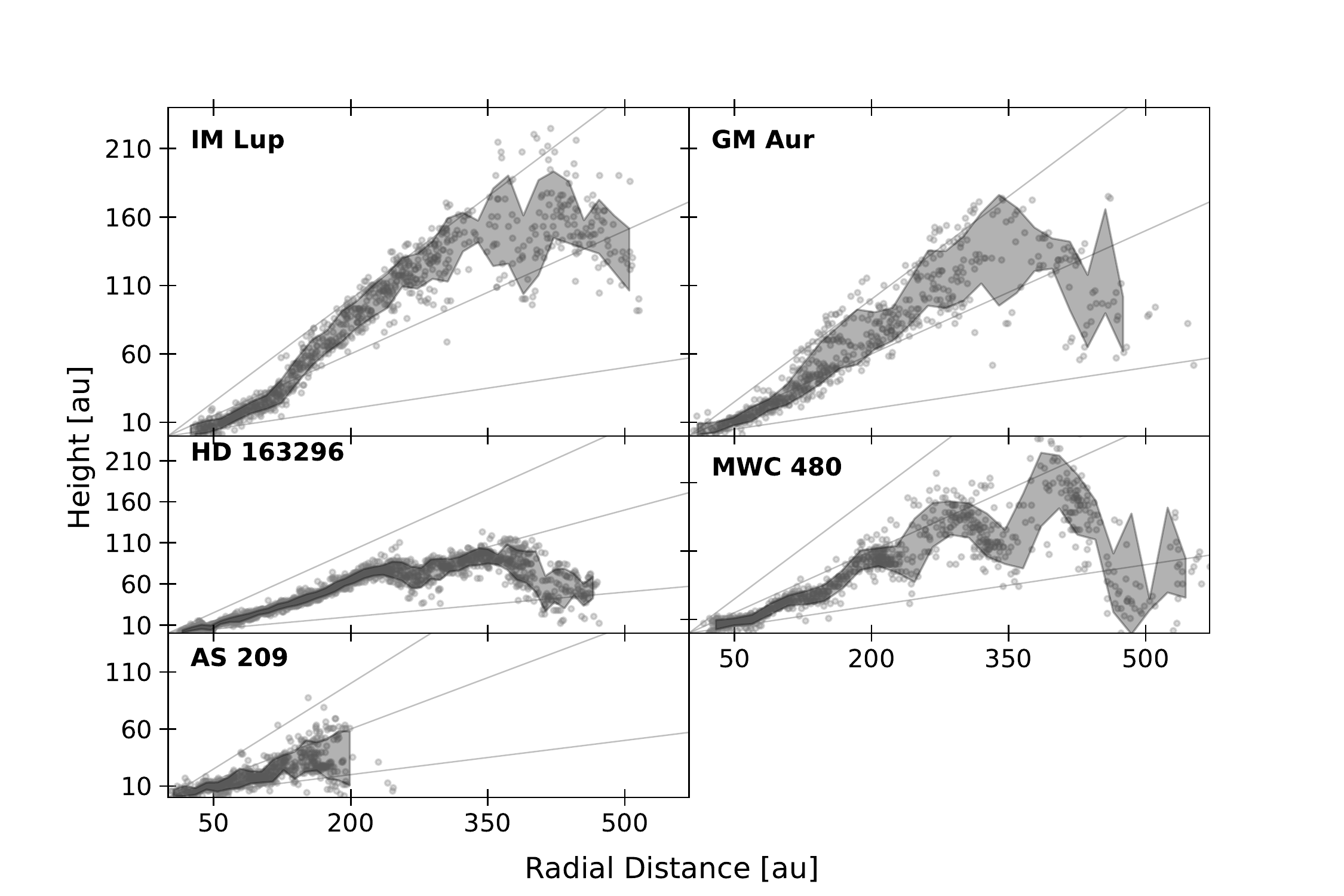}
      \caption{$^{12}$CO $J = 2-1$ data points
              }
         \label{panel_12CO_dots}
\end{figure*}

\begin{figure*}[h!]
   \centering
   \includegraphics[scale=0.65]{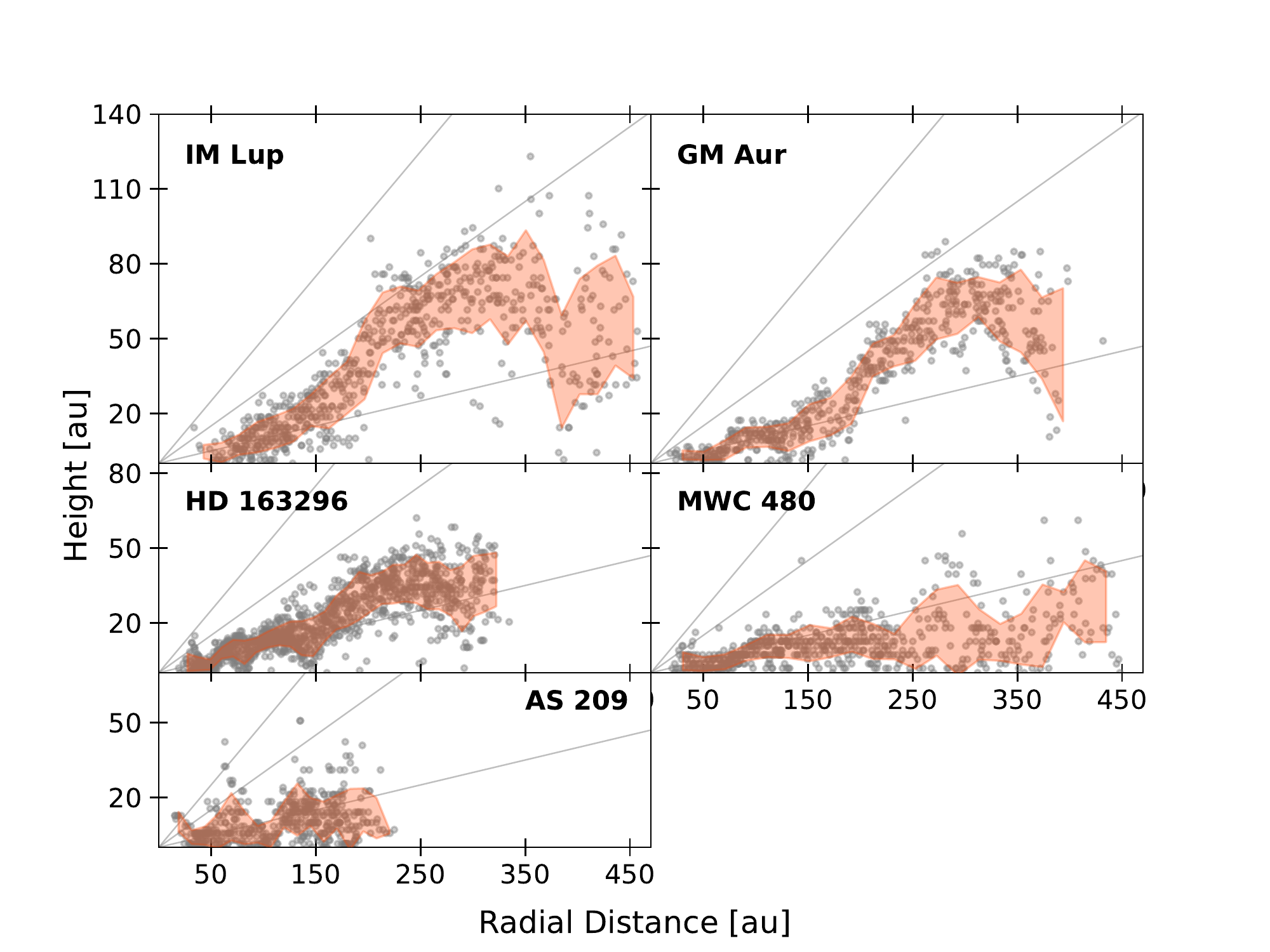}
      \caption{$^{13}$CO $J = 2-1$ data points
              }
         \label{panel_13CO_dots}
\end{figure*}

\begin{figure*}[h!]
   \centering
   \includegraphics[scale=0.65]{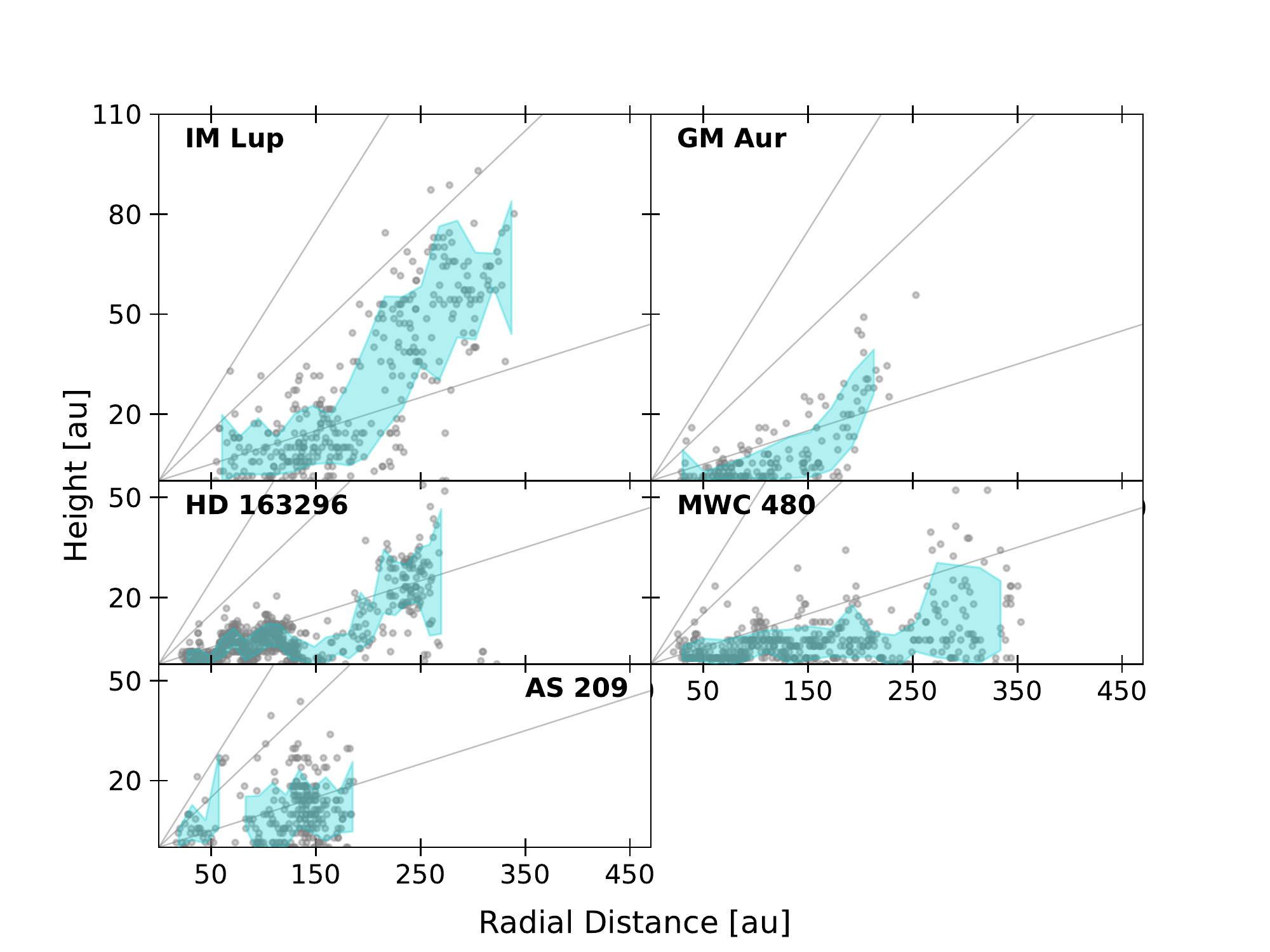}
      \caption{C$^{18}$O $J = 2-1$ data points
              }
         \label{panel_C18O_dots}
\end{figure*}

\begin{figure*}[h!]
   \centering
   \includegraphics[scale=0.65]{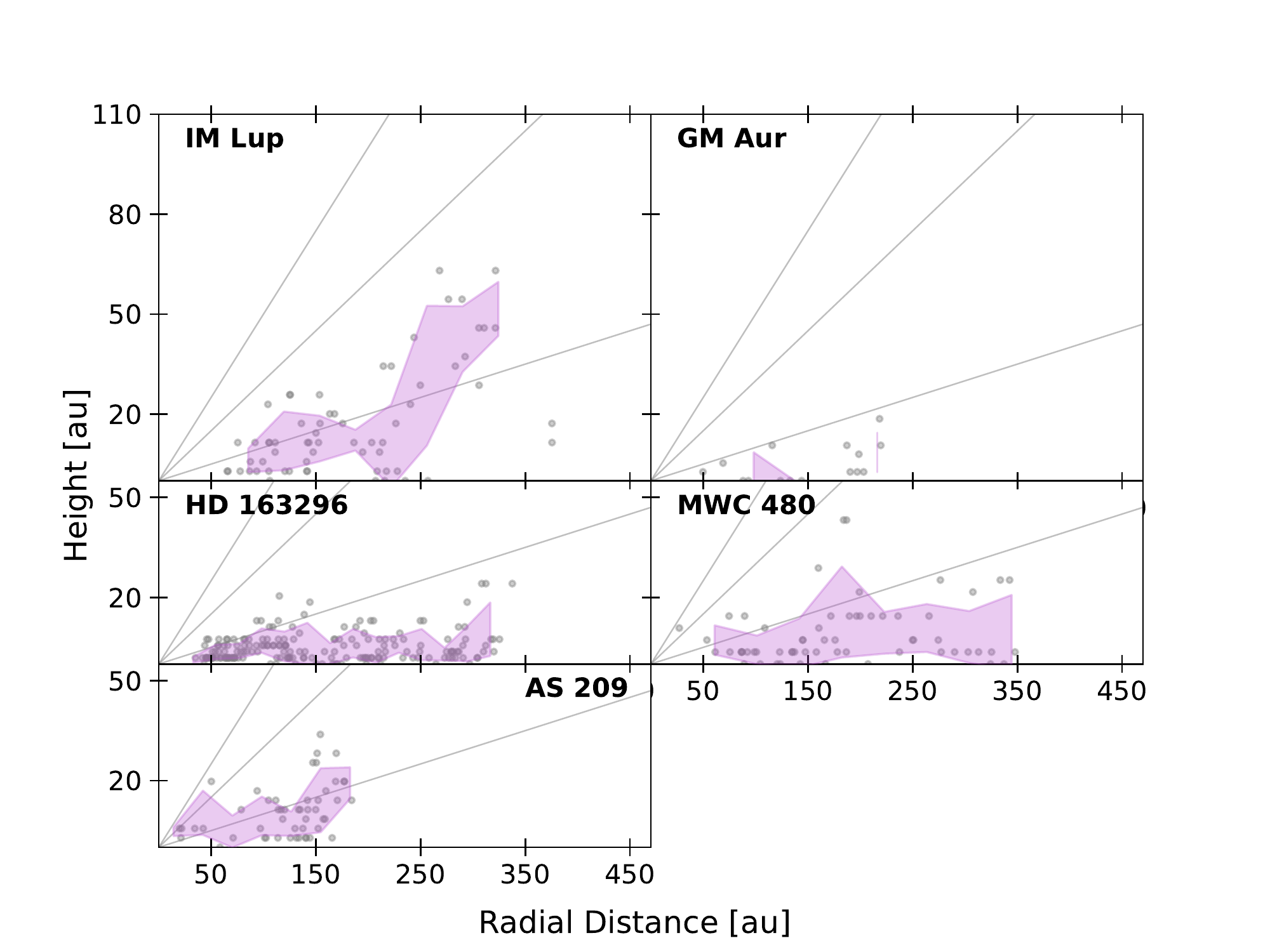}
      \caption{$^{13}$CO $J=1-0$ data points
              }
         \label{panel_13CO_2_dots}
\end{figure*}

\begin{figure*}[h!]
   \centering
   \includegraphics[scale=0.65]{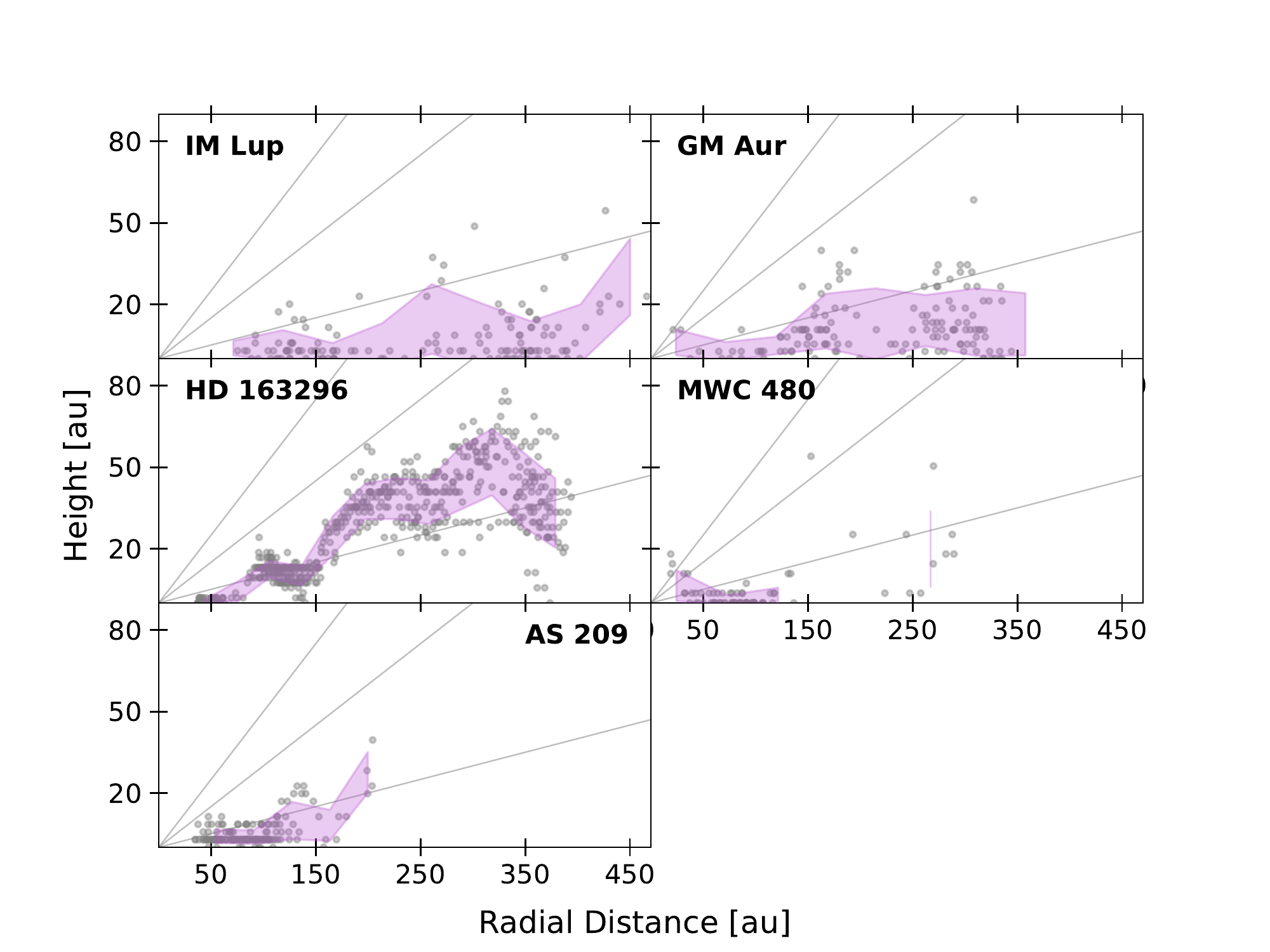}
      \caption{HCN data points
              }
         \label{panel_HCN_dots}
\end{figure*}

\begin{figure*}[h!]
   \centering
   \includegraphics[scale=0.65]{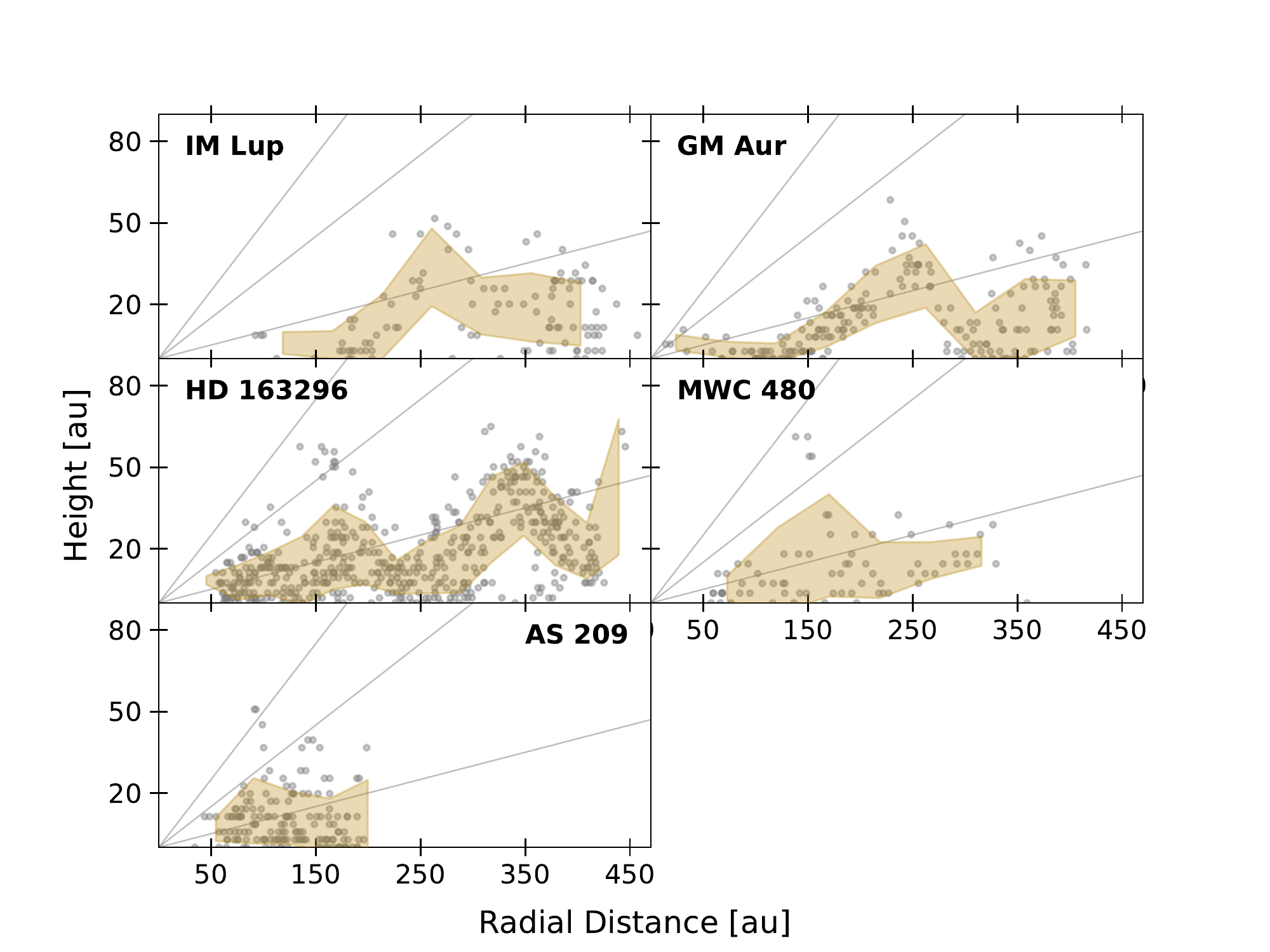}
      \caption{H$_2$CO data points
              }
         \label{panel_H2CO_dots}
\end{figure*}

\begin{figure*}[h!]
   \centering
   \includegraphics[scale=0.65]{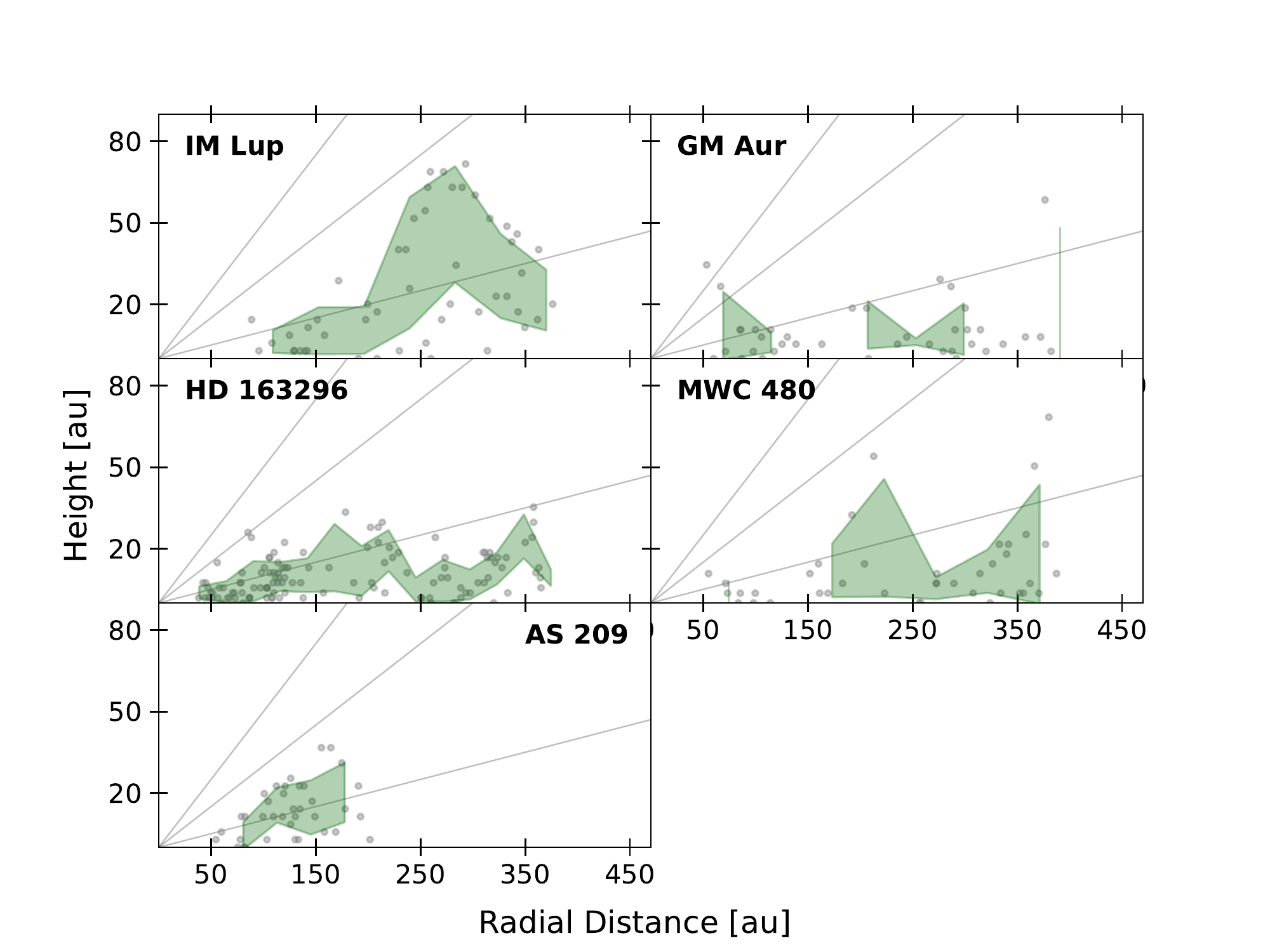}
      \caption{HCO$^+$ data points
              }
         \label{panel_HCO_dots}
\end{figure*}

\begin{figure*}[h!]
   \centering
   \includegraphics[scale=0.65]{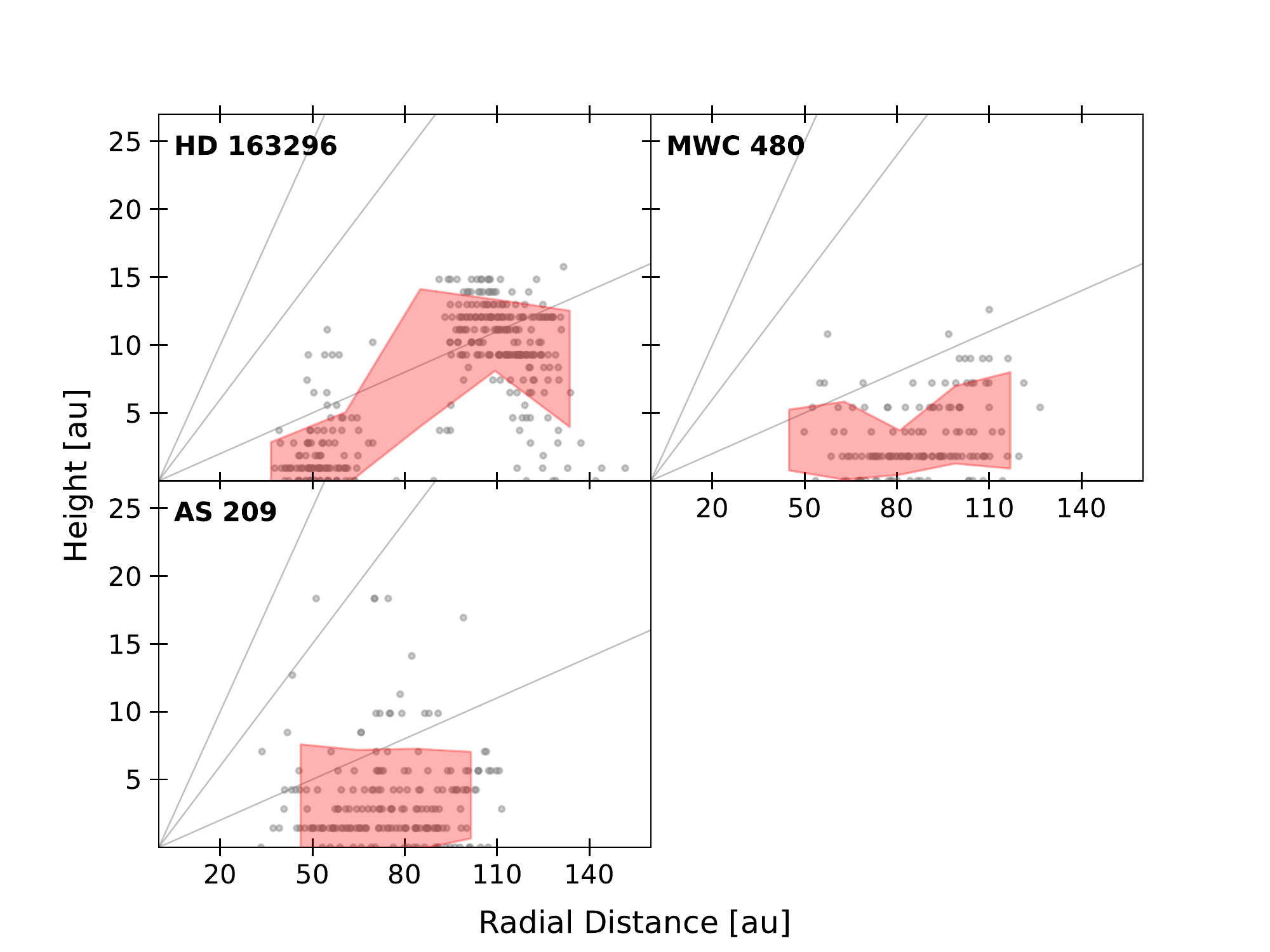}
      \caption{C$_2$H data points
              }
         \label{panel_C2H_dots}
\end{figure*}

\begin{figure*}[h!]
   \centering
   \includegraphics[width=\hsize]{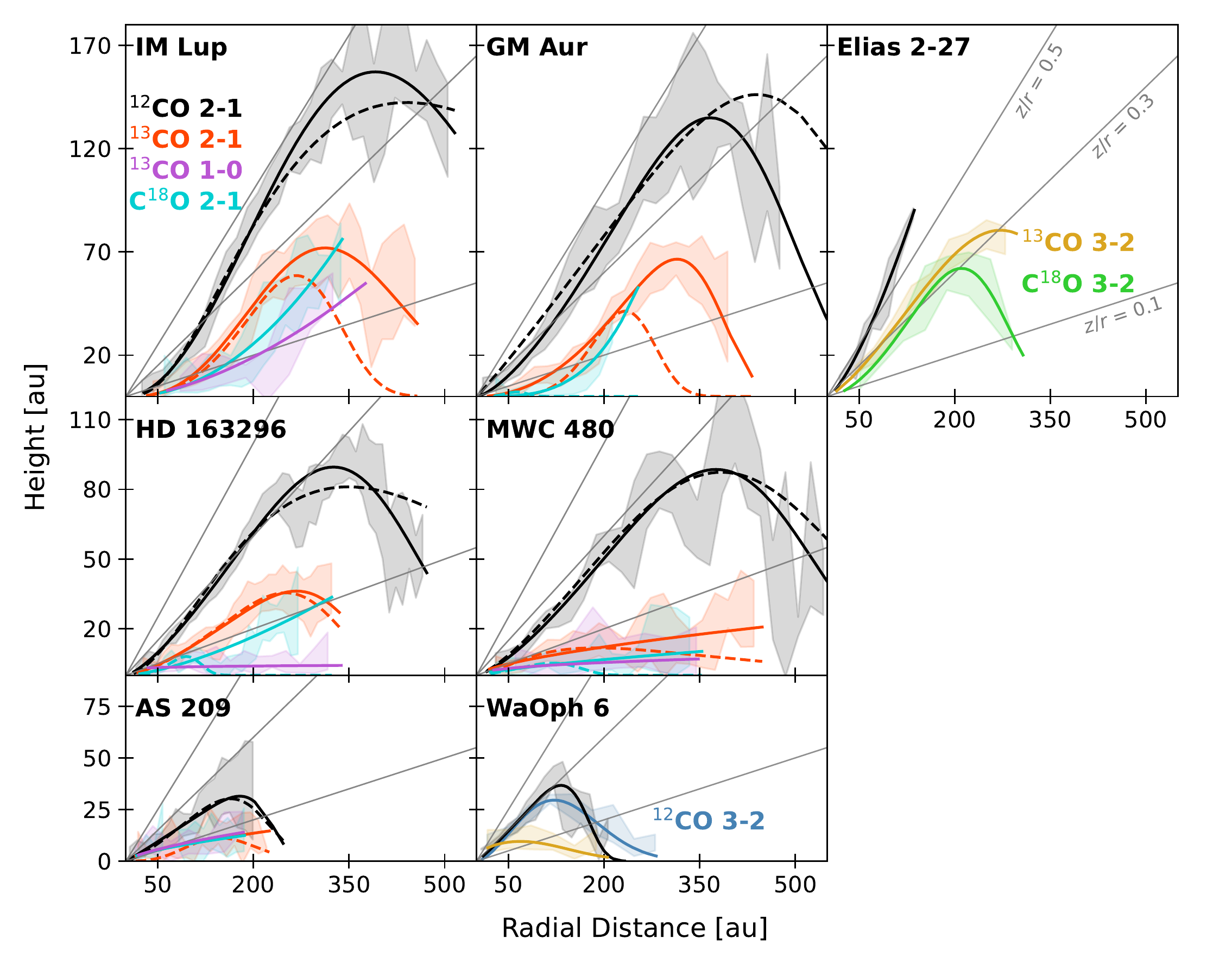}
      \caption{Best-fit vertical profiles of CO isotopologues in solid colored lines, compared when possible to previously derived emission surfaces by Law et al. 2020b in dashed lines. Shaded background colors show the dispersion of the retrieved data. Solid grey lines indicate the curves of constant $z/r$ for values of 0.1, 0.3 and 0.5.
              }
         \label{panel_CO_bestfit_diskminer}
\end{figure*}

For each molecule and disk in the MAPS sample we present the extracted emission surfaces, showing all of the obtained data points in Figures \ref{panel_12CO_dots}, \ref{panel_13CO_dots},  \ref{panel_C18O_dots},  \ref{panel_13CO_2_dots},  \ref{panel_HCN_dots},  \ref{panel_H2CO_dots},  \ref{panel_HCO_dots} and  \ref{panel_C2H_dots}. The averaged vertical profiles and the best-fit parameters for each CO transition profile are shown in Figure \ref{panel_CO_bestfit_diskminer} compared to those found using DISKSURF \citep{disksurf} in \citet{MAPS_Law_Surf}. In general the surfaces are similar for $^{12}$CO, however in $^{13}$CO and C$^{18}$O the previous work underestimates the height of the emission compared to our results. This is particularly apparent beyond 200\,au. The best-fit values for the emitting surfaces of all CO isotopologues, parameterized using an exponentially tapered power-law can be found in Table \ref{table_vertical_co}.

\begin{table*}[h]
\def\arraystretch{1.5}
\setlength{\tabcolsep}{5pt}
\caption{Best-fit parameters of vertical profiles for CO isotopologues}
\label{table_vertical_co}      
\centering
\begin{tabular}{c c c c c c}       
\hline\hline                
Source & Line & $z_0$ (au) & $\phi$ &  $r_{taper}$ (au) & $\psi$
 \\    
\hline \hline                       
    
IM Lup &    $^{12}$CO 2$-$1 & 25.99 $^{+0.29}_{-0.25}$ & 1.95 $^{+0.04}_{-0.03}$ & 417.65 $^{+10.26}_{-11.58}$ & 2.26 $^{+0.1}_{-0.11}$  \\ 
        &   $^{13}$CO 2$-$1 & 9.76 $^{+0.25}_{-0.27}$ & 2.65 $^{+0.12}_{-0.1}$ & 310.32 $^{+12.65}_{-14.28}$ & 2.6 $^{+0.16}_{-0.18}$  \\
        &   C$^{18}$O 2$-$1 & 6.07 $^{+0.18}_{-0.25}$ & 2.07 $^{+0.04}_{-0.03}$ & - & -  \\
        &   $^{13}$CO 1$-$0 & 5.97 $^{+0.5}_{-0.48}$ & 1.67 $^{+0.09}_{-0.1}$ & - & -  \\ 

\hline
GM Aur &    $^{12}$CO 2$-$1 & 27.54 $^{+0.18}_{-0.22}$ & 1.45 $^{+0.02}_{-0.01}$ & 470.29 $^{+2.38}_{-2.02}$ & 4.94 $^{+0.24}_{-0.26}$  \\   
        &   $^{13}$CO 2$-$1 & 7.99 $^{+0.23}_{-0.22}$ & 2.11 $^{+0.04}_{-0.04}$ & 372.93 $^{+1.49}_{-1.44}$ & 7.19 $^{+0.36}_{-0.38}$  \\ 
        &   C$^{18}$O 2$-$1 & 2.25 $^{+0.28}_{-0.24}$ & 3.4 $^{+0.16}_{-0.17}$ & - & -  \\  
        &   $^{13}$CO 1$-$0 & 0.08 $^{+0.5}_{-0.06}$ & 6.19 $^{+1.85}_{-2.52}$ & - & -  \\

\hline 

Elias 2-27 &    $^{12}$CO 2$-$1 & 58.65 $^{+0.39}_{-0.52}$ & 1.38 $^{+0.04}_{-0.03}$ & - & -  \\ 
        &   $^{13}$CO 3$-$2 & 31.69 $^{+0.48}_{-0.34}$ & 1.3 $^{+0.04}_{-0.03}$ & 360.79 $^{+7.7}_{-10.04}$ & 3.57 $^{+0.65}_{-0.54}$  \\
        &   C$^{18}$O 3$-$2 & 25.28 $^{+0.41}_{-0.34}$ & 1.68 $^{+0.06}_{-0.06}$ & 262.34 $^{+3.41}_{-4.16}$ & 4.75 $^{+0.33}_{-0.31}$  \\ 

\hline
HD 163296 &    $^{12}$CO 2$-$1 & 25.08 $^{+0.12}_{-0.11}$ & 1.35 $^{+0.01}_{-0.01}$ & 426.44 $^{+1.35}_{-1.27}$ & 4.21 $^{+0.09}_{-0.1}$  \\ 
        &   $^{13}$CO 2$-$1 & 11.88 $^{+0.14}_{-0.14}$ & 1.4 $^{+0.03}_{-0.02}$ & 343.88 $^{+4.79}_{-3.86}$ & 5.36 $^{+0.4}_{-0.44}$  \\
        &   C$^{18}$O 2$-$1 & 6.02 $^{+0.2}_{-0.14}$ & 1.47 $^{+0.04}_{-0.03}$ & - & -  \\ 
        &   $^{13}$CO 1$-$0 & 3.65 $^{+0.17}_{-0.23}$ & 0.13 $^{+0.14}_{-0.09}$ & - & -  \\ 
\hline        
MWC 480 &    $^{12}$CO 2$-$1 &  19.82 $^{+0.17}_{-0.17}$ & 1.38 $^{+0.02}_{-0.02}$ & 489.74 $^{+4.33}_{-4.62}$ & 4.25 $^{+0.24}_{-0.22}$  \\
        &   $^{13}$CO 2$-$1 & 7.72 $^{+0.13}_{-0.13}$ & 0.66 $^{+0.02}_{-0.02}$ & - & -  \\ 
        &   C$^{18}$O 2$-$1 & 4.08 $^{+0.17}_{-0.14}$ & 0.73 $^{+0.05}_{-0.05}$ & - & -  \\ 
        &   $^{13}$CO 1$-$0 & 4.04 $^{+0.46}_{-0.35}$ & 0.44 $^{+0.16}_{-0.22}$ & - & -  \\
\hline      
AS 209 &    $^{12}$CO 2$-$1 & 19.49 $^{+0.26}_{-0.2}$ & 1.05 $^{+0.03}_{-0.03}$ & 229.98 $^{+2.5}_{-2.13}$ & 8.12 $^{+0.76}_{-0.7}$  \\  
        &   $^{13}$CO 2$-$1 & 8.54 $^{+0.17}_{-0.18}$ & 0.66 $^{+0.04}_{-0.05}$ & - & -  \\
        &   C$^{18}$O 2$-$1 & 8.36 $^{+0.34}_{-0.29}$ & 0.64 $^{+0.1}_{-0.08}$ & - & -  \\
        &   $^{13}$CO 1$-$0 & 9.17 $^{+0.57}_{-1.15}$ & 0.69 $^{+0.33}_{-0.15}$ & - & -  \\
        
\hline
WaOph 6 &    $^{12}$CO 3$-$2 &  44.82 $^{+9.37}_{-5.39}$ & 1.6 $^{+0.22}_{-0.2}$ & 140.52 $^{+13.07}_{-16.47}$ & 2.18 $^{+0.28}_{-0.27}$  \\
        &   $^{12}$CO 2$-$1 & 31.69 $^{+0.69}_{-0.64}$ & 1.12 $^{+0.03}_{-0.03}$ & 173.07 $^{+2.9}_{-1.58}$ & 6.67 $^{+0.99}_{-1.47}$  \\
        &   $^{13}$CO 3$-$2 & 13.24 $^{+9.46}_{-2.12}$ & 0.41 $^{+0.33}_{-0.2}$ & 145.19 $^{+43.16}_{-33.62}$ & 2.34 $^{+4.64}_{-1.07}$  \\

\hline \hline                                   
\end{tabular}
\end{table*}

\begin{table*}[h]
\def\arraystretch{1.6}
\setlength{\tabcolsep}{5pt}
\caption{Best-fit parameters of vertical profiles for other tracers in HD\,163296.}
\label{table_vertical_hd16}      
\centering
\begin{tabular}{c c c c c c}       
\hline\hline                
Molecule & Transition & $z_0$ (au) & $\phi$ &  $r_{taper}$ (au) & $\psi$ \\
\hline\hline
CN &$N = 1-0$ & 10.92 $^{+0.37}_{-0.41}$ & 1.62 $^{+0.05}_{-0.05}$ & 401.11 $^{+2.81}_{-2.98}$ & 7.54 $^{+0.38}_{-0.36}$  \\
HCN &$J = 3-2$ & 8.45 $^{+0.15}_{-0.12}$ & 2.14 $^{+0.07}_{-0.08}$ & 342.11 $^{+8.07}_{-9.06}$ & 3.75 $^{+0.38}_{-0.35}$  \\
H$_2$CO &$J = 3_{03} - 2_{02}$& 19.2 $^{+1.66}_{-2.41}$ & 1.0 $^{+0.09}_{-0.06}$ & 462.85 $^{+331.83}_{-302.32}$ & 0.07 $^{+0.09}_{-0.05}$  \\ 
C$_2$H &$N = 3-2$ & 38.13 $^{+27.14}_{-12.6}$ & 5.41 $^{+0.74}_{-0.72}$ & 95.38 $^{+9.65}_{-11.73}$ & 3.78 $^{+0.81}_{-0.74}$  \\
HCO$^{+}$ & $J = 1-0$ & 6.45 $^{+0.41}_{-0.46}$ & 0.59 $^{+0.07}_{-0.08}$ & - & -  \\
c-C$_3$H$_2$ &$J = 7_{07} - 6_{16}$ & 9.56 $^{+0.57}_{-0.64}$ & 2.79 $^{+0.27}_{-0.28}$ & - & -  \\ 
\hline \hline

\end{tabular}
\tablefoot{These values are used to deproject the emission and extract the brightness temperature profiles presented in section 3.1 }
\end{table*}
\subsection{Modulations in vertical surface}

\begin{figure*}[h!]
   \centering
   \includegraphics[width=\hsize]{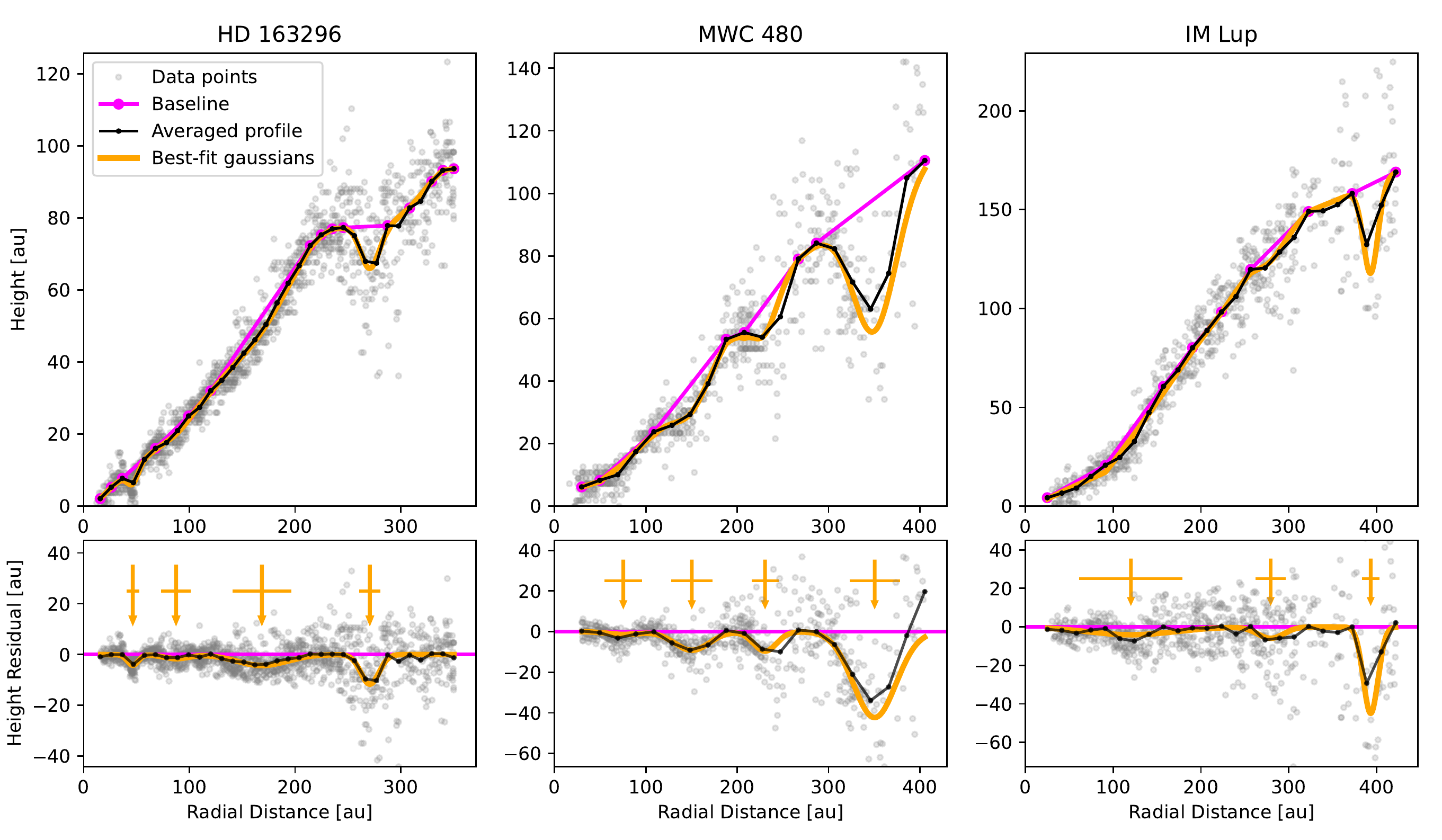}
      \caption{This panel displays the data (grey dots), averaged profile (black curve), assumed baseline (pink profile) and best-fit gaussians (orange line) obtained for each system under analysis in section 3.2.1 for $^{12}$CO. The bottom panels show the residuals obtained after subtracting the baseline profile to the data points. The vertical arrows mark the best-fit center of each gaussian component considered in each system. The horizontal line marks the width of the gaussian component.
              }
         \label{gauss_vert_12CO}
\end{figure*}

\begin{figure*}[h!]
   \centering
   \includegraphics[width=\hsize]{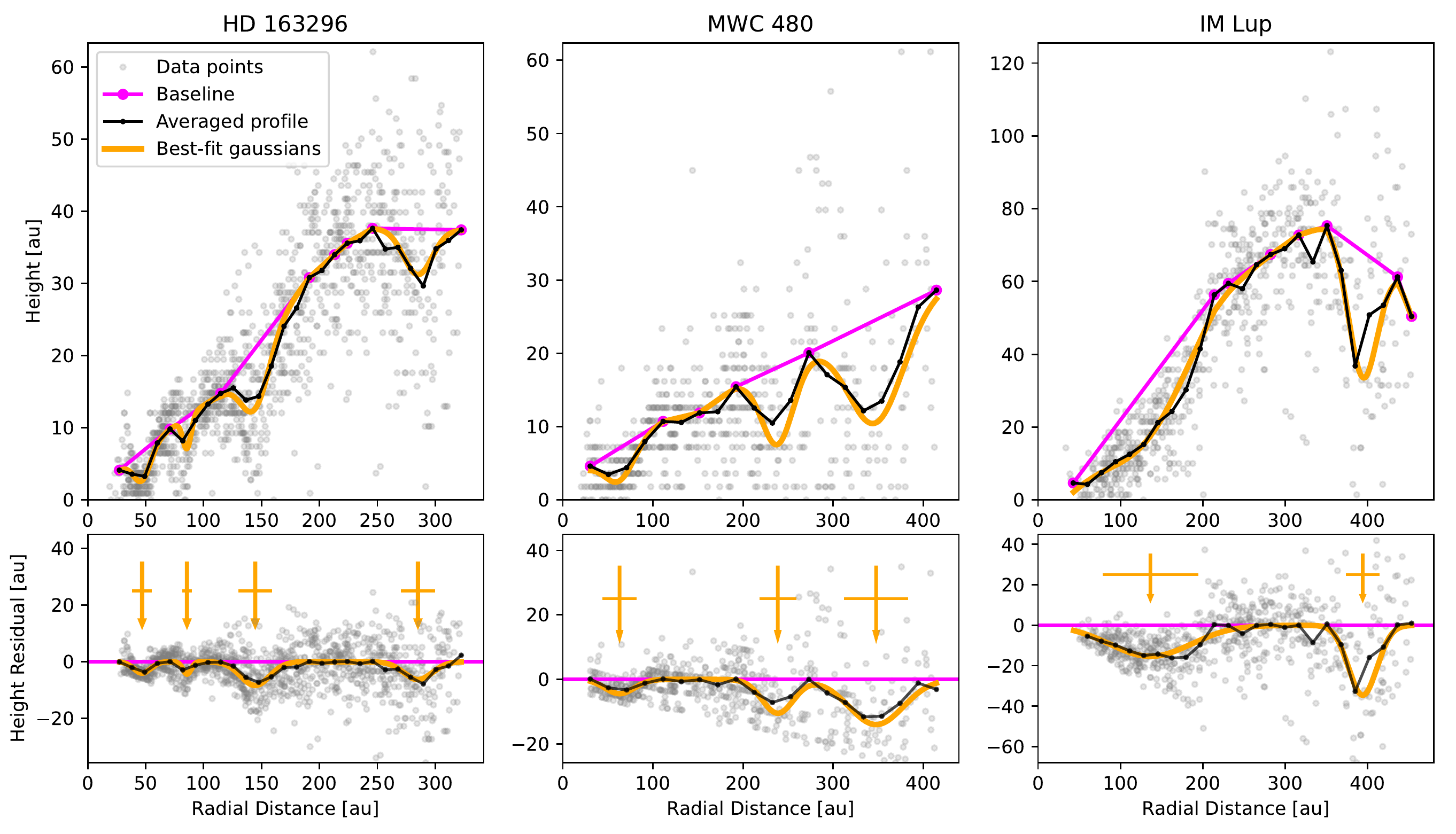}
      \caption{Same as Figure \ref{gauss_vert_12CO} but for $^{13}$CO.
              }
         \label{gauss_vert_13CO}
\end{figure*}

\begin{figure*}[h!]
   \centering
   \includegraphics[width=\hsize]{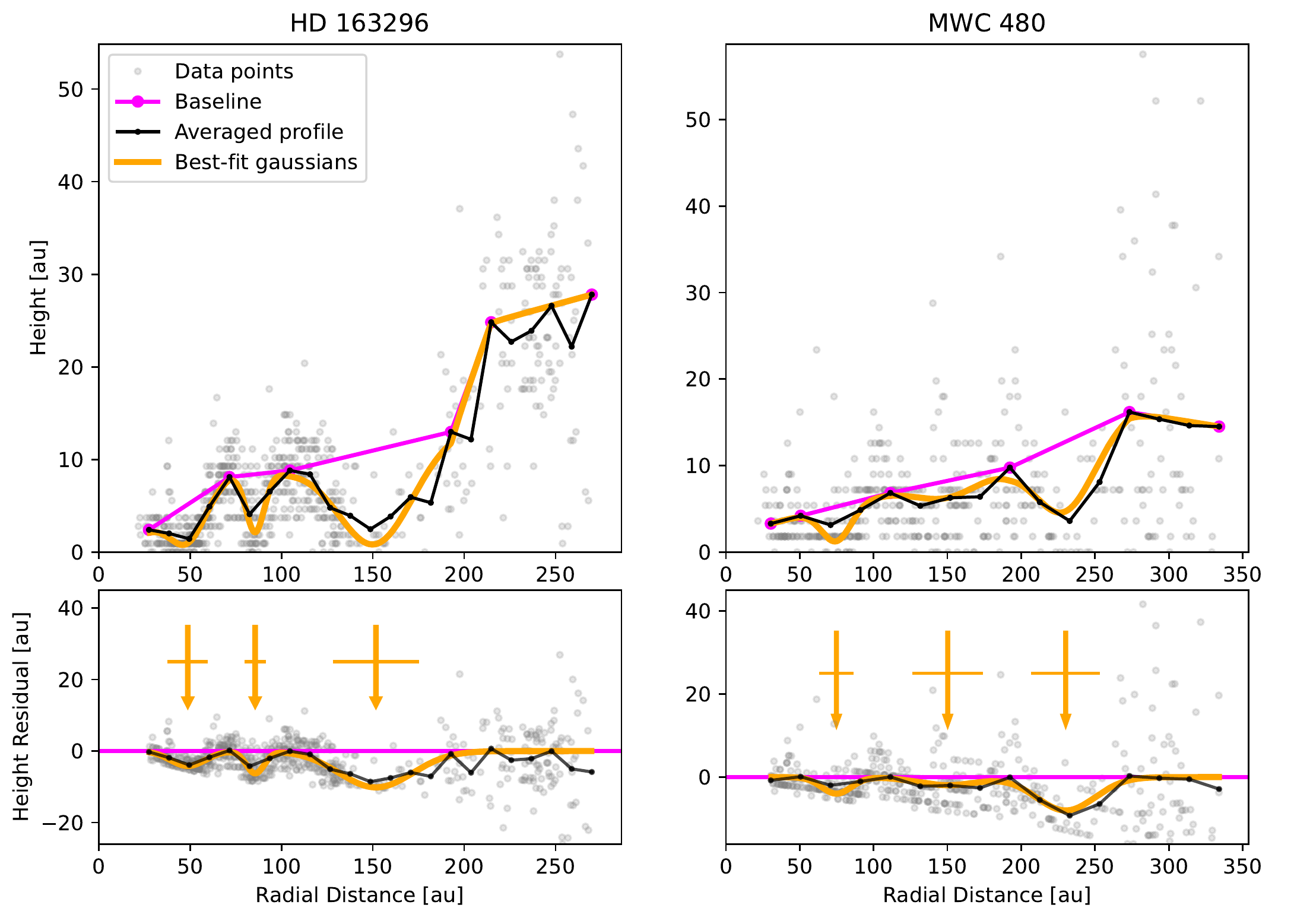}
      \caption{Same as Figure \ref{gauss_vert_12CO} but for C$^{18}$O.
              }
         \label{gauss_vert_C18O}
\end{figure*}

The modulations detected in the CO isotopologues for IM Lup, HD 163296 and MWC 480 are characterized as detailed in section 3.2.1. Figures \ref{gauss_vert_12CO}, \ref{gauss_vert_13CO} and \ref{gauss_vert_C18O} display the baselines, averaged data and location of best-fit gaussians for each disk and tracer analyzed. The values of the best-fit location, width and depth of the vertical gaussian components for each studied disk and isotopologue are shown in Table \ref{table_gauss_vert}.

\begin{table*}[h]
\def\arraystretch{1.5}
\setlength{\tabcolsep}{5pt}
\caption{Modulations in CO vertical surface for HD\,163296, MWC\,480 and IM\,Lup.}
\label{table_gauss_vert}      
\centering
\begin{tabular}{c c c c c c}       
\hline\hline                
Source & Line & $r_0$ (au) & Width (au) &  $\Delta z$ (au) & Depth ($\Delta z/z(r_0)$)
 \\    
\hline \hline                       
          
   HD\,163296 & $^{12}$CO 2$-$1 &46.5$\pm$0.51 & 11.96$\pm$2.05 & 4.21 & 0.41 \\  
            &               & 87.42$\pm$4.04 & 28.21$\pm$8.92 & 1.64 & 0.08 \\ 
            &               & 168.61$\pm$2.36 & 55.64$\pm$5.75 & 4.33 & 0.08 \\ 
            &               & 270.76$\pm$1.22 & 20.28$\pm$3.18 & 11.66 & 0.15 \\    
\hline             
            & $^{13}$CO 2$-$1 & 46.94$\pm$0.67 & 17.04$\pm$1.53 & 4.01 & 0.6 \\ 
            &               & 85.68$\pm$0.9 & 8.45$\pm$2.03 & 4.36 & 0.38 \\
            &               & 144.54$\pm$0.81 & 29.45$\pm$2.12 & 8.38 & 0.4 \\ 
            &               & 285.01$\pm$1.89 & 29.16$\pm$3.6 & 6.31 & 0.17 \\   
\hline             
            & C$^{18}$O 2$-$1 & 48.77$\pm$0.53 & 22.13$\pm$1.3 & 4.19 & 0.81 \\ 
            &               & 85.63$\pm$0.57 & 11.5$\pm$1.04 & 6.22 & 0.74 \\   
            &               & 151.7$\pm$2.24 & 47.17$\pm$3.86 & 10.17 & 0.92 \\ 
\hline\hline 
   MWC\,480 & $^{12}$CO 2$-$1 & 75.24$\pm$3.71 & 41.41$\pm$13.82 & 1.37 & 0.09 \\ 
            &               & 150.23$\pm$1.7 & 44.95$\pm$4.1 & 9.19 & 0.23 \\   
            &               & 230.48$\pm$1.63 & 29.52$\pm$4.45 & 9.62 & 0.15 \\   
            &               & 350.48$\pm$2.18 & 55.1$\pm$3.06 & 42.27 & 0.43 \\   
\hline              
             & $^{13}$CO 2$-$1 & 62.89$\pm$1.52 & 38.34$\pm$3.08 & 4.46 & 0.63 \\ 
             &               & 238.47$\pm$1.74 & 40.88$\pm$4.38 & 10.51 & 0.58 \\
             &               & 347.62$\pm$2.62 & 70.83$\pm$5.25 & 14.05 & 0.57 \\
\hline             
            & C$^{18}$O 2$-$1 & 74.83$\pm$0.8 & 23.51$\pm$2.17 & 3.99 & 0.76 \\ 
            &               & 150.18$\pm$1.08 & 47.9$\pm$1.74 & 1.96 & 0.24 \\  
            &               & 230.13$\pm$1.09 & 46.67$\pm$2.28 & 8.05 & 0.63 \\ 
\hline\hline 
   IM\,Lup & $^{12}$CO 2$-$1 & 120.12$\pm$4.4 & 117.51$\pm$14.44 & 4.12 & 0.11 \\   
            &               & 279.49$\pm$1.89 & 34.75$\pm$5.25 & 5.93 & 0.05 \\ 
            &               & 393.49$\pm$0.88 & 19.66$\pm$2.78 & 44.92 & 0.28 \\
\hline             
            & $^{13}$CO 2$-$1 & 136.34$\pm$2.1 & 116.29$\pm$1.58 & 15.35 & 0.47 \\
            &               & 393.97$\pm$0.9 & 40.59$\pm$2.24 & 34.55 & 0.51 \\

\hline \hline                                   
\end{tabular}
\end{table*}

\section{Tests and comparisons of the gas pressure scale height}

\subsection{Comparison scattering surfaces to CO and gas pressure scale height.}

\begin{figure}[h!]
   \centering
   \includegraphics[width=\hsize]{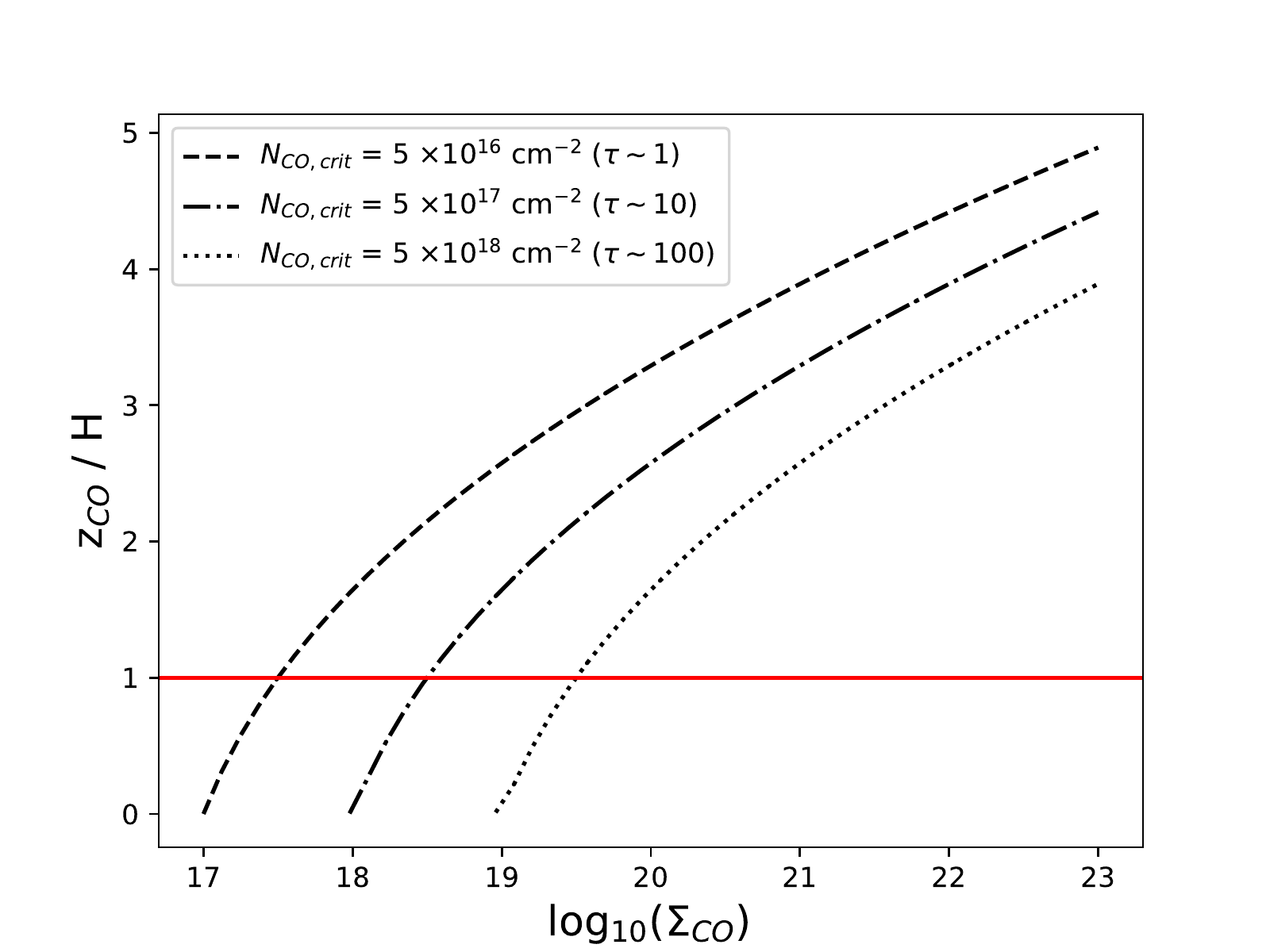}
      \caption{Conversion factor between the $^{12}$CO emitting surface ($z_{\mathrm{CO}}$) and the gas pressure scale height ($H$) as a function of the assumed CO density profile. This result is obtained by solving equation 5 and varying the critical density value for CO as indicated in the legend.
              }
         \label{relation_hydro_dens}
\end{figure}

From our simple one layer model detailed in section 3.2.3 we relate the location of the CO emission layer to that of the gas pressure scale height. This relation has a strong dependence on the surface density profile, CO abundance and critical density, that varies according to the optical depth ($\tau$) at the emission location. Figure \ref{relation_hydro_dens} shows the relation between these values. 

\begin{figure}[h!]
   \centering
   \includegraphics[width=\hsize]{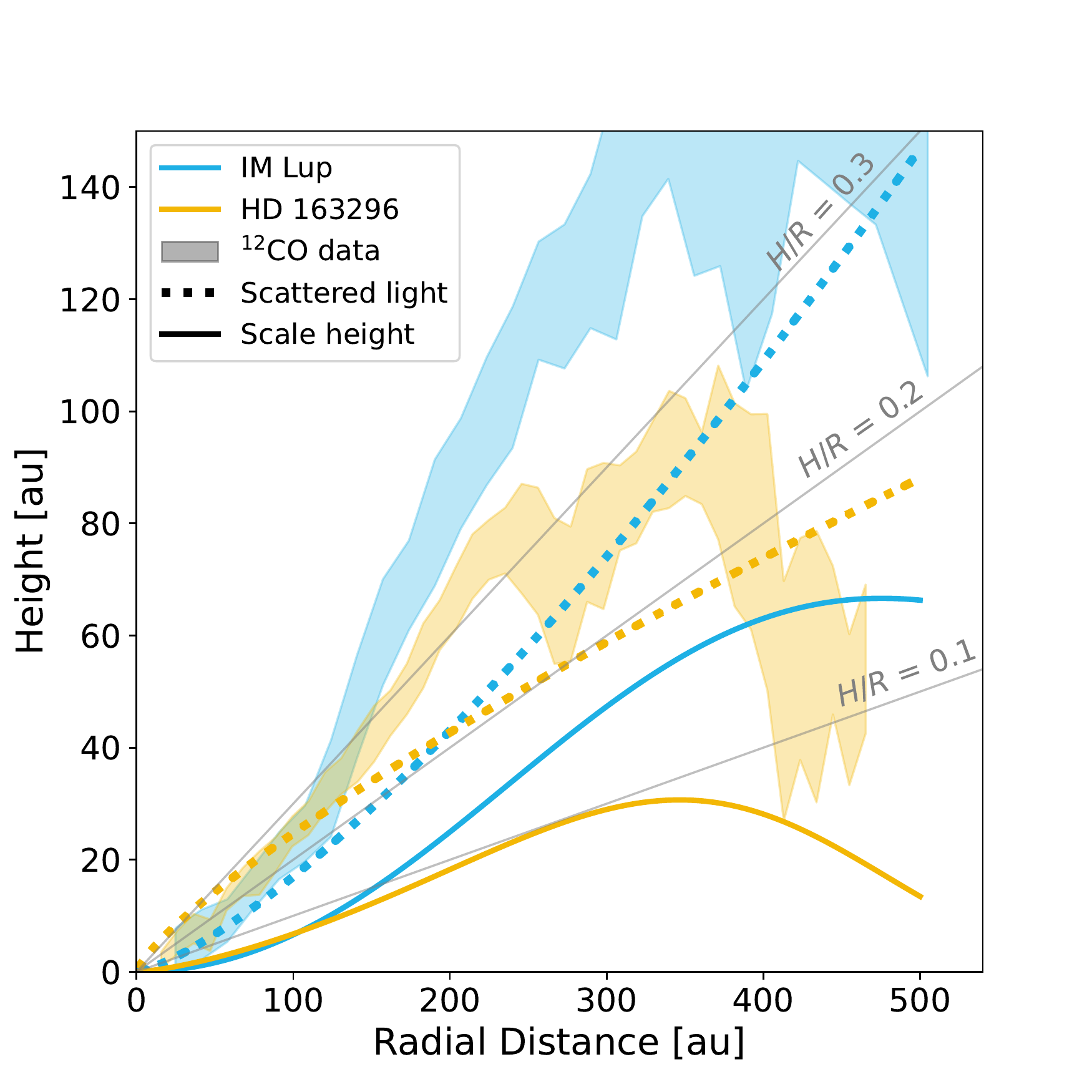}
      \caption{For IM Lup in blue and HD 163296 in yellow, broad shaded region shows the location of the emission surface of $^{12}$CO $J=2-1$ as extracted in this work. Dotted lines indicate the scattering surfaces as parametrized in \citet{Rich_2021} and solid colored lines indicate the parametrization of the gas pressure scale height derived in this work. 
              }
         \label{comp_hd16_imlup_height}
\end{figure}

Figure \ref{comp_hd16_imlup_height} shows the vertical location of the $^{12}$CO $J=2-1$ emission layer, the parametrization of the scattering surface \citep{Rich_2021} and the inferred location of the gas pressure scale height obtained in this work. It is seen that for both IM Lup and HD 163296 the $^{12}$CO layer traces the most vertically extended region, followed by the scattering surface and closer to the midplane lies the gas pressure scale height. 

\subsection{Conversion factors between $z_{\mathrm{CO}}$ and $H$ and alternative estimate }

\begin{figure}[h!]
   \centering
   \includegraphics[scale=0.45]{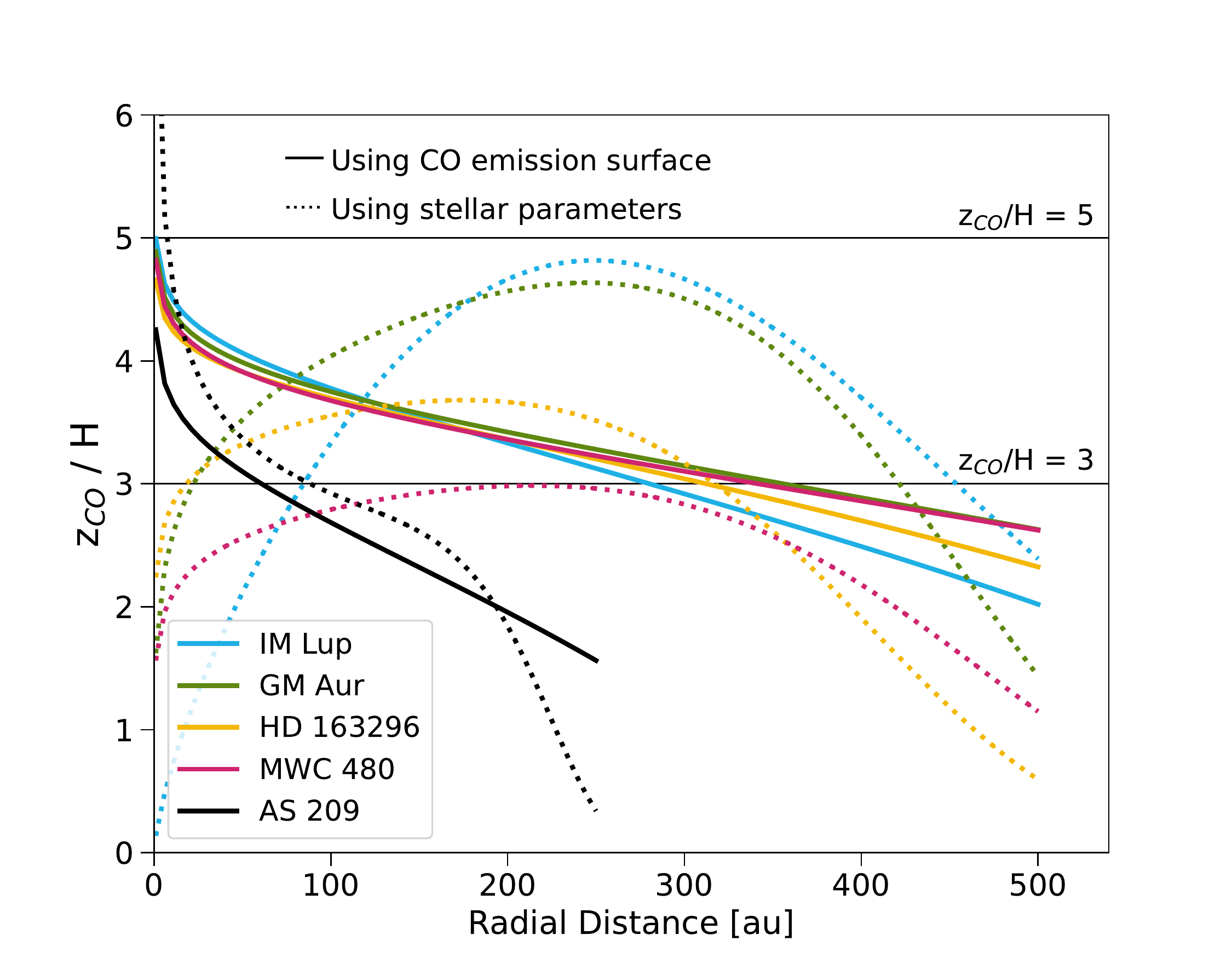}
      \caption{Comparison of conversion factor between $^{12}$CO emitting surface ($z_{\mathrm{CO}}$) and gas pressure scale height ($H$) obtained through our simple single layer model (solid line) and using an estimate on the midplane temperature from the stellar parameters (dotted line).
              }
         \label{hydro_scale_comp}
\end{figure}

Another method to estimate the pressure scale height is through the midplane temperature of the disk, calculated considering the luminosity of the star and then related to the scale height through the stellar mass \citep[see equations 3 and 4 from ][]{Law_2022_12CO}. This has been done for several disks with estimates on their CO emission layers, including the MAPS sample in \citet{Law_2022_12CO}. Figure \ref{hydro_scale_comp} shows the comparison of the radial conversion factors between $z_{\mathrm{CO}}$ and $H$ obtained through our single layer model and what is obtained from the pressure scale height derived through stellar parameters, assuming a flaring angle of 0.02. The conversion factors stretch between 3-5 in the inner 300\,au and then to lower values in the outer radii for all of our sample. The curves shown in Figure \ref{hydro_scale_comp} are smoother than those derived in \citet{Law_2022_12CO} because we are using the best-fit parametrization of the CO emitting layer instead of the average height values. We note that, with the exception of AS\,209, our method obtains a much narrower range for the conversion factor than when considering the stellar parameters. Through our test with thermochemical models we have confirmed the accuracy of our method, which implies that the method using stellar parameters is not a good approximation to the pressure scale height. The narrow range of our result suggests that there may be a direct relation, varying with radius, to convert the location of the CO emitting layer into gas pressure height. The difference in the conversion profile of AS\,209 is due to the much lower density it has been reported to have \citep{MAPS_Zhang}. Overall, these high resolution and sensitivity data open new avenues in our understanding of the disk physical structure.

\subsection{DALI models setup}

An initial test on our simple analytical method to estimate the gas pressure scale height is done through thermochemical DALI \citep{Bruderer_DALI_2013, Bruderer_2012} models. For details on the code, how the chemical network is set and the calculation of the temperature structure, dust populations and flux values we reference to past works using the same code \citep[e.g.][]{leemker_2022_LkCa15, miotello_2016, Bruderer_2012, Bruderer_DALI_2013}. We use the chemical network of \citet{Bruderer_2012} and consider a chemical age of 1\,Myr for the calculations. Other general parameters common to all simulations are presented in Table \ref{table_DALI_param}.

\begin{table}
\caption{Common parameters in DALI models}            
\label{table_DALI_param}   
\def\arraystretch{1.3}
\centering                         
\begin{tabular}{l l}       
\hline\hline               
Parameter & Value \\   
\hline       
    $Chemistry$ & \\
    Chemical Age & 1\,Myr \\
   {[CO]}/{[H]}  & 1.35 $\times$ 10$^{-4}$ \\
   {[PAH]} &  10$^{-4}$ ISM abundance\\
\hline
   $Physical$ $Structure$ & \\
   gas-to-dust ratio & 100 \\      
   $f_\mathrm{{large}}$ (fraction small/large grains) & 0.85  \\
   $\chi$ (Settling of large grains) & 0.2 \\ 
\hline            
    $Stellar$ $Parameters$ & \\
    $L_x$ & 1.0 $\times$ 10$^{30}$ erg\,s$^{-1}$\\
    $T_x$ & 7.0 $\times$ 10$^{7}$ K \\
    $\zeta_\mathrm{{c.r.}}$ (cosmic ray ionization rate) & 5.0 $\times$ 10$^{-17}$ s$^{-1}$\\
\hline
\end{tabular}
\end{table}

The aim of these models is to create a structure using the best-fit gas surface density values reported in \citet{MAPS_Zhang} for each of the MAPS sources \citep[see Table 2 in ][]{MAPS_Zhang}. The radial structure of the gas and dust in DALI follows the self-similar solution to a viscously evolving disk \citep{Lynden_Bell_Pringle_1974, Andrews_2011}.

\begin{equation}
    \Sigma(R) = \Sigma_c \left( \frac{R}{R_c} \right) ^{-\gamma} \mathrm{exp} \left[ - \left(\frac{R}{R_c} \right) ^{2-\gamma} \right]
\end{equation}

where $R_c$ is the characteristic radius and $\gamma$ the surface density exponent. The vertical structure of the disk follows a Gaussian distribution, which we have modified from the typical DALI prescription \citep[see for example][]{leemker_2022_LkCa15} to match the prescription of \citet{MAPS_Zhang}. The scale height angle ($h$) is set at each radius following,

\begin{equation}
    h = h_{100} \left( \frac{R}{ 100 \mathrm{au} } \right) ^{\varphi}
\end{equation}

where $h_{100}$ is the scale height angle at 100\,au. The scale height angle can be converted to a physical height above the disk midplane, $H$ as $H \sim hR$. We note that as the scale height is defined in angular units, the flaring angle $\varphi$ will be $ 1 - \varphi_r$, where $\varphi_r$ is the flaring angle for a model in physical units \citep[as presented in ][]{MAPS_Zhang}. The physical and stellar parameters used for each disk are shown in Table \ref{table_DALI_disks}. Using the radiative transfer tools of DALI \citep{Bruderer_DALI_2013, Bruderer_2012}, we create mock channel maps from each model matching the velocity (0.2\,km\,s$^{-1}$) and spatial (0.13\arcsec) resolution of the $^{12}$CO observations. These channel maps are analyzed using ALFAHOR in the same way as the observations (see section 2.2.1) to extract the emitting layer.

\begin{figure}[h!]
   \centering
   \includegraphics[width=\hsize]{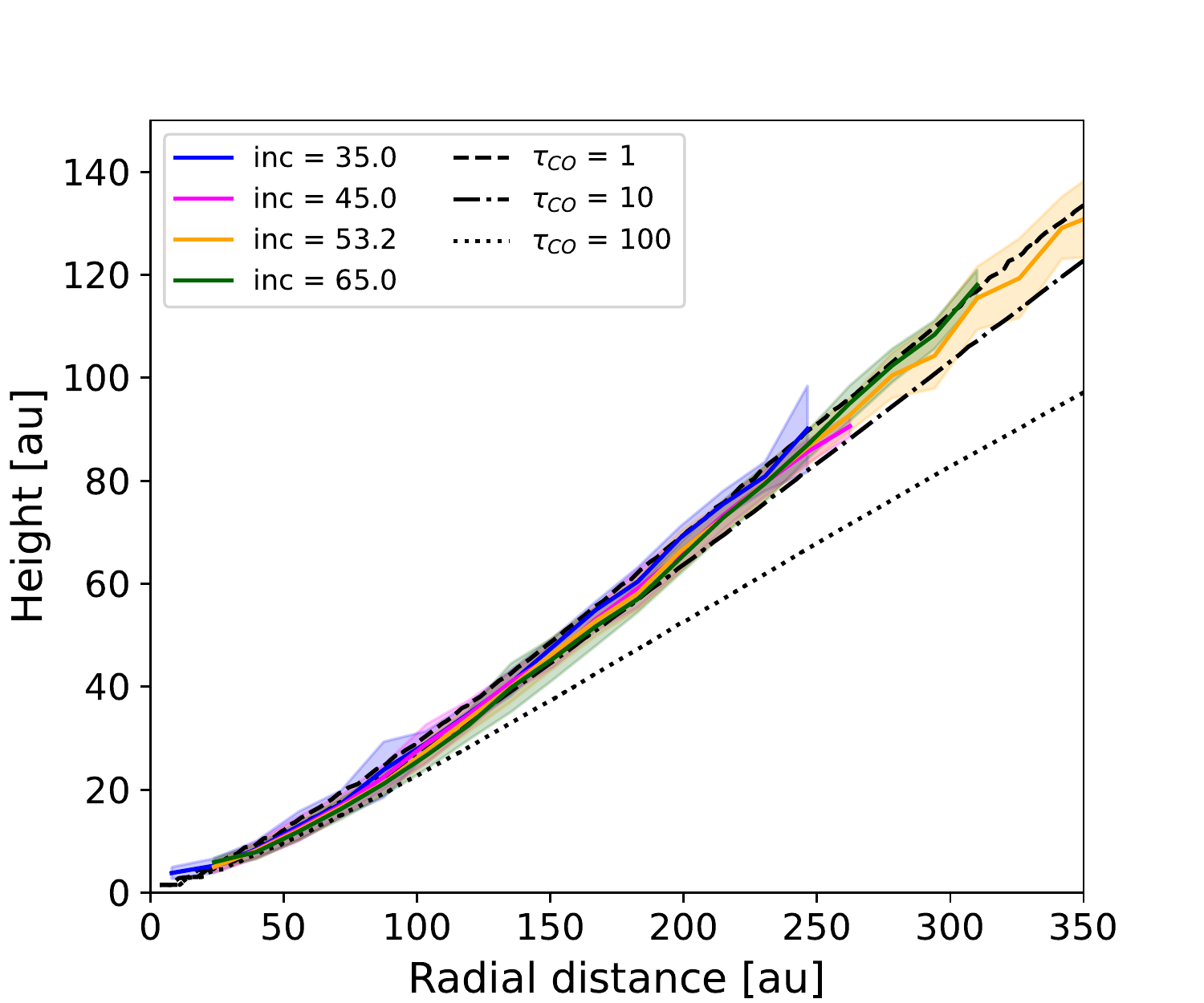}
      \caption{Extracted vertical profiles for $^{12}$CO $J=2-1$ obtained using mock channel maps at different inclinations, from the GM\,Aur DALI model. Different colors trace the various inclinations and the black lines indicate the $^{12}$CO millimeter optical depth values, as done in Figure \ref{DALI_test}.
              }
         \label{DALI_inc}
\end{figure}

Using the DALI models we additionally test if the inclination of the system may affect on our retrieval of the vertical surface. Figure \ref{DALI_inc} shows the retrieved vertical profiles for $^{12}$CO $J=2-1$ from the GM\,Aur model using three different inclinations to compute the channel maps. As can be seen, there is no significant variation between the inclinations, indicating that, at least for optically thick tracers, the orientation of the disk is not a source of concern. What may vary at different inclinations is the radial extent up to which it is possible to confidently separate and trace far and near surfaces in the channels. This leads to a difference in the maximum radial region in which we can trace the vertical structure, but, within a radial extent that is sampled by all inclinations, the vertical profiles are in agreement. Further testing with the models of the other studied systems show the same results of coherent values at different inclinations. IM\,Lup is the disk that shows largest variations, however the values are still very similar within the uncertainties. As discussed in section 3.2.3 the surface density model of IM\,Lup considers the most massive and extended disk of the sample and this could be related to tracing a lower vertical layer and slight differences with varying inclinations. This will be studied in detail in future work focused on thermochemical models.

\begin{table*}
\caption{Physical parameters for DALI models}            
\label{table_DALI_disks}   
\def\arraystretch{1.3}
\centering                         
\begin{tabular}{c c c c c c c c | c c c c}       
\hline\hline               
Source & $\Sigma_c$ & $R_c$ & $\gamma$ & $h_{\mathrm{100}}$ & $\varphi$ & $r_{\mathrm{in}}$ & $r_{\mathrm{out}}$ & $T_{eff}$ & $L_*$ & $M_*$ & log$_{10}(\dot{M})$ \\ 
 & (g\,cm$^{-2}$) & (au) & & (rad) & & (au) & (au) & (K) & (L$_{\odot}$) & (M$_{\odot}$) & (M$_{\odot}$\,yr$^{-1}$) \\
\hline       

IM\,Lup & 28.4 & 100 & 1.0 & 0.1 & 0.17 & 0.20 & 1200 & 4266 & 2.57 & 1.1 & -7.9   \\
GM\,Aur & 9.4 & 176 & 1.0 & 0.075 & 0.35 & 1.00 & 650 & 4350 & 1.2 & 1.1 & -8.1   \\
AS\,209 & 1.0 & 80 & 1.0 & 0.06 & 0.25 & 0.5 & 300 & 4266 & 1.41 & 1.2 & -7.3  \\
HD\,163296 & 8.8 & 165 & 0.8 & 0.084 & 0.08 & 0.45 & 600 & 9332 & 17.0 & 2.0 & -7.4 \\
MWC\,480 & 5.8 & 200 & 1.0 & 0.1 & 0.08 & 0.45 & 750 & 8250 & 21.9 & 2.1 & -6.9  \\

\hline
\end{tabular}
\tablefoot{Values are based on the stellar parameters and best-fit gas surface density results presented in \citet{MAPS_Zhang}.
    }
\end{table*}

\end{appendix}

\end{document}